\documentclass[11pt]{article}
\usepackage[T1]{fontenc}
\usepackage{lmodern}
\usepackage[letterpaper,margin=1in]{geometry}
\usepackage{graphicx}
\usepackage{xcolor}
\usepackage{booktabs}
\usepackage{amsmath,amssymb}
\usepackage{microtype}
\usepackage[numbers,sort&compress]{natbib}
\usepackage{setspace}
\usepackage{hyperref}
\usepackage{bm}
\hypersetup{hidelinks}
\graphicspath{{figures/}}
\newcommand{\cspbi}{CsPbI$_3$}
\newcommand{\ASPH}{ASPH}

\begin{document}

\title{Chemically Resolved Topological Coordinates Link Structural Dynamics and Configurational Thermodynamics}
\author{Ayush Kumar Pandey$^{1*}$ and Abhishek Tewari$^{2,3*}$\\[4pt]
\small $^1$Department of Physics, Indian Institute of Technology Roorkee, Uttarakhand, India\\
\small $^2$Department of Metallurgical and Materials Engineering, Indian Institute of Technology Roorkee,\\
\small Uttarakhand, India\\
\small $^3$Mehta Family School of Data Science and Artificial Intelligence,\\
\small Indian Institute of Technology Roorkee, Uttarakhand, India\\[3pt]
\small $^*$Corresponding authors: \texttt{ayush\_p@ph.iitr.ac.in}; \texttt{abhishek@mt.iitr.ac.in}}
\date{}

\maketitle

\begin{abstract}
Atomic coordinates specify a structure, but they do not reveal how chemical connectivity across several length scales relates to atomic motion and configurational energy ordering. We formulate chemically directed persistent homology at four resolutions---complete networks, individual sites, spatial fields, and substitutional arrangements---while retaining the chemical identity and length scale of each connectivity feature. In \textit{ab initio} molecular dynamics (AIMD) trajectories of $\delta$- and $\gamma$-\cspbi\ at five temperatures spanning 500--700 K, the corner-sharing $\gamma$ network has a lower Pb--I restoring stiffness and permits larger iodide excursions, yet iodide positional correlations decay 2.27 times more slowly and Pb-network topology retains memory 1.63 times longer than in the edge-sharing $\delta$ phase. Local softness and loss of network memory are therefore distinct. At individual sites, rare $\gamma$-phase Pb environments with a $\delta$-like Cs-cage connectivity precede 0.17~\AA\ greater Pb displacement over the subsequent 0.5 ps. The same Pb-network coordinate resolves disruption of corner-sharing connectivity across a 2560-atom $\delta|\gamma$ boundary. In substituted \cspbi, compact dopant arrangements undergo greater cooperative host relaxation and lie lower in density-functional-theory (DFT) energy than dispersed arrangements of the same composition. SchNet and Allegro model families with comparable energy errors encode opposite ordering along this coordinate, and 61 of 247 supplied models with errors below 1 meV atom$^{-1}$ on separate test structures reverse the DFT relation. Because relative configurational energies set Boltzmann populations, chemical network topology links structure to physical response and tests whether learned energy models preserve DFT configurational ordering even when their average errors are small.
\end{abstract}

\section{Introduction}

Atomic coordinates specify a structure, but they do not by themselves identify which chemical organization is relevant to its subsequent dynamics or thermodynamic selection. Distances, angles, coordination numbers, and symmetry modes provide direct physical coordinates when the relevant local motif is known and remains intact. When bonds and coordination networks reorganize, however, an intermediate or interfacial environment may not possess the ideal motif used to define such a coordinate. A complementary description should retain chemical identity and connectivity beyond the first coordination shell, remain invariant to rigid translation, rotation, and permutations of identical atoms, and preserve its meaning from a collective network to an individual site or spatially heterogeneous region.

Crystal chemistry provides an extensive set of interpretable structural measures. Bond lengths and angles, octahedral tilts and distortions, and symmetry-adapted modes directly connect atomic motion to a known coordination unit. PDynA systematizes such time-dependent octahedral descriptors for metal-halide perovskites,\cite{Liang2023PDynA} while Pyrovskite supplies construction, analysis, and featurization tools for two- and three-dimensional perovskite structures.\cite{Stanton2023Pyrovskite} Bond-orientational order parameters instead summarize local angular symmetry without naming a particular polyhedron.\cite{Steinhardt1983} These measures are concise because they answer a specific structural question. An octahedral coordinate requires six neighbors that can be assigned to one octahedron, while a local bond or angle does not state how successive coordination units connect over longer distances.

General-purpose representations relax these assumptions. The smooth overlap of atomic positions (SOAP) expands a smooth neighbor density into a rotation-invariant, systematically improvable basis and is widely used both as a structural similarity measure and as an input to regression.\cite{Bartok2013SOAP,Himanen2020DScribe} Its scalar meaning, however, is task dependent: a high-dimensional SOAP vector must generally be projected or fitted before it becomes a reaction coordinate or an interfacial field. Graph neural networks learn still more flexible representations and can reproduce atomistic energies and forces with high accuracy.\cite{Schutt2018SchNet,Musaelian2023Allegro} Energy and force errors measure predictive accuracy, but they do not reveal whether two configurations at the same composition are ordered for the correct structural reason. A complementary, low-dimensional coordinate is therefore useful both for interpreting trajectories and for testing a learned energy landscape.

Persistent homology follows the connectivity of a point cloud as the connection distance grows and records the scales over which components, loops, and enclosed arrangements persist. The resulting filtration is invariant to rigid motion and to permutations of identical points, and it resolves multiple length scales within a stated outer neighborhood instead of assigning one bond cutoff. Persistent homology has provided direct physical insight into medium-range organization in amorphous networks\cite{Hiraoka2016,Sorensen2020} and ion-migration pathways.\cite{Sato2023} For crystalline materials, ASPH has been used to predict formation energies,\cite{Jiang2021ASPH} and local persistent-homology vectors have been added to graph networks for defect-energy prediction.\cite{Fang2025Defects} Path-based topology and other recent extensions broaden the structural relations that can be represented, while a current review surveys their mathematical foundations and materials applications.\cite{Chen2023PathTopology,Zheng2026Review} In a multicomponent crystal, a single unlabeled point cloud combines sublattices with different chemical roles. ASPH preserves those roles through separate filtrations for chosen center and neighbor species.\cite{Jiang2021ASPH} In halide perovskites, topological representations have distinguished compositions and finite-temperature phases and have predicted band gaps;\cite{Anand2022Perovskite} quotient-complex descriptors have been applied to two-dimensional perovskite design;\cite{Hu2025Quotient} cohomological distances have quantified similarity among perovskite configurations;\cite{Wee2025Cohomology} and ASPH has represented layered hybrid lead halides for band-gap learning.\cite{Marchenko2025Layered} Here the chemical channel, homology dimension, and filtration scale are instead retained as named physical coordinates and compared directly with independent dynamical and thermodynamic quantities. Complete-network, site-resolved, spatially resolved, and substitutional forms of the same chemical filtration then test whether its structural meaning persists across resolution.

The polymorphism of \cspbi\ provides such a test. Its black perovskite phases contain a three-dimensional network of corner-sharing PbI$_6$ octahedra, whereas yellow $\delta$-\cspbi\ contains one-dimensional double chains of edge-sharing octahedra. The $\delta\leftrightarrow\gamma$ transformation must therefore rebuild Pb--I connectivity rather than simply condense a single small-amplitude mode; calculations resolve a multistep pathway with three intermediate structures.\cite{Li2022Pathway} Differential scanning calorimetry places the reversible phase change between 548 and 599 K,\cite{Dastidar2017} while a later free-energy calculation combining RPA+HF ground-state energies with machine-learned vibrational contributions predicts 500 K.\cite{Braeckevelt2022} The difference illustrates the sensitivity of the transition temperature to the electronic-structure and free-energy treatment. Strong anharmonicity, spatially correlated octahedral motion, and transient lower-symmetry regions further broaden the finite-temperature structural landscape.\cite{Baldwin2024} More broadly, computational studies identify ion migration and structural instability as coupled sources of halide-perovskite degradation,\cite{Bhatt2023Review} while composition screening shows that band-gap and vacancy-energy targets must be considered together.\cite{Pandey2026Mixed} The same network question appears in two technologically relevant settings: where the polymorphs coexist, the boundary contains environments that need not resemble either ideal bulk crystal; under Cd or Zn substitution, both dopant arrangement and host-network relaxation contribute to stability.

\begin{figure}[!htbp]
\centering
\includegraphics[width=\textwidth]{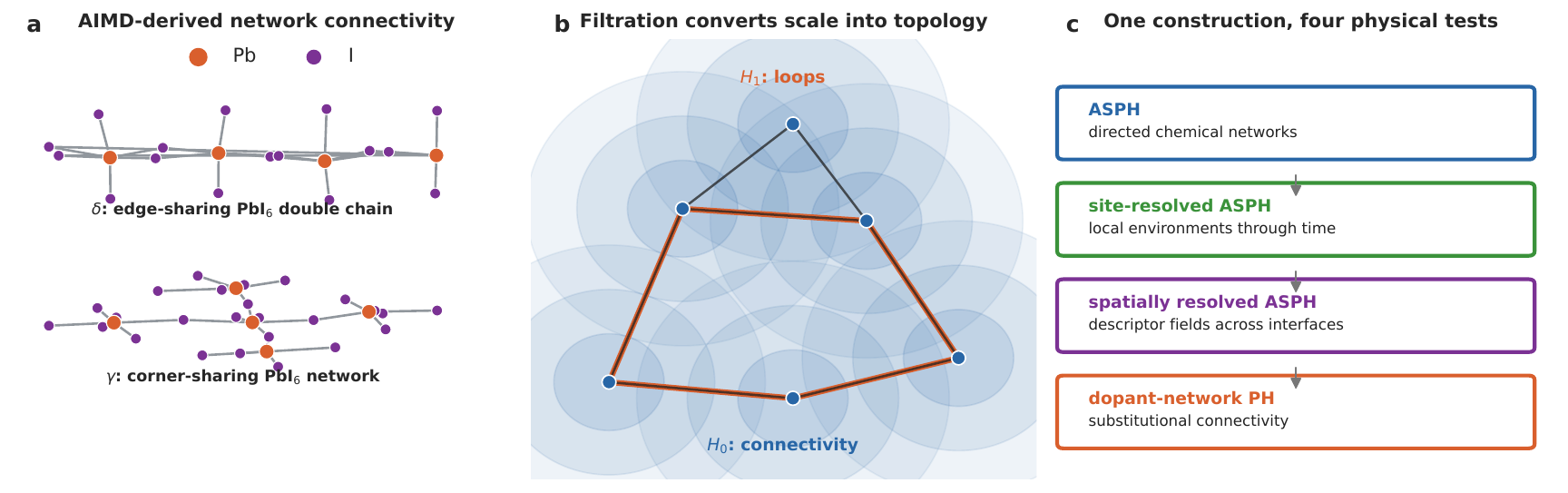}
\caption{\label{fig:concept}One chemical-network construction, four physical tests. (a) Clusters from ab initio molecular-dynamics (AIMD) trajectories show the edge-sharing $\delta$ double chain and corner-sharing $\gamma$ network. (b) Growing connection radii illustrate when separate groups join ($H_0$) and loops form ($H_1$); $H_2$, used in later analyses, records enclosed three-dimensional features. (c) Complete species networks are related to structural memory and relaxation; atom-centered environments to later displacement; a spatially resolved network field to interfacial disruption; and substitutional networks to energy ordering and configurational populations.}
\end{figure}

We test whether chemically resolved network topology can serve as the same physical coordinate at four levels of atomic organization: a fluctuating bulk network, an individual atomic environment, a structurally mixed phase boundary, and a substitutional configuration space (Figure~\ref{fig:concept}). Each chemical direction, homology dimension, and filtration scale is retained as a named structural quantity instead of being pooled only into a predictive vector. The test is whether these quantities remain linked to independently calculated responses: network memory and iodide relaxation in the bulk, subsequent Pb displacement at a site, loss of corner-sharing connectivity at a boundary, and relaxation energy and configurational population under substitution. The final comparison asks whether graph-neural-network energies preserve the DFT ordering of arrangements along the same dopant-connectivity coordinate. In this way, persistent homology is used to measure chemical network organization rather than merely to separate structural classes.

\section{Results and Discussion}

\subsection{Network connectivity separates local softness from structural memory}

The edge-sharing $\delta$ chains and the three-dimensional corner-sharing $\gamma$ network remain topologically distinct throughout the 500--700 K trajectories. Each phase was simulated in a 320-atom cell at 500, 550, 600, 650, and 700 K, with 10 ps of equilibration followed by 64 ps of production at every state point. To test whether the phase differences persist in time, each production trajectory was divided into five successive portions; atoms and individual frames within a portion were averaged before phases were compared (SI Sections S1--S2 and Tables S1--S2).

For a central atom of species $A$, the directed $A\rightarrow B$ construction contains that atom and surrounding atoms of species $B$. Increasing the connection distance reveals when initially separate groups join ($H_0$), when loops appear and close ($H_1$), and when enclosed three-dimensional features appear and fill ($H_2$). The direction identifies the chemical viewpoint: I$\rightarrow$Pb describes the Pb framework surrounding an iodide, whereas Pb$\rightarrow$Cs describes the Cs cage surrounding a framework Pb. The three possible central species and three surrounding species give nine directed chemical channels. The 9.5 \AA\ neighborhood includes several coordination shells and remains within the unique periodic minimum-image range of every analyzed frame; the radius test is reported in SI Section S3 and Table S4.

\begin{figure}[!htbp]
\centering
\includegraphics[width=\textwidth]{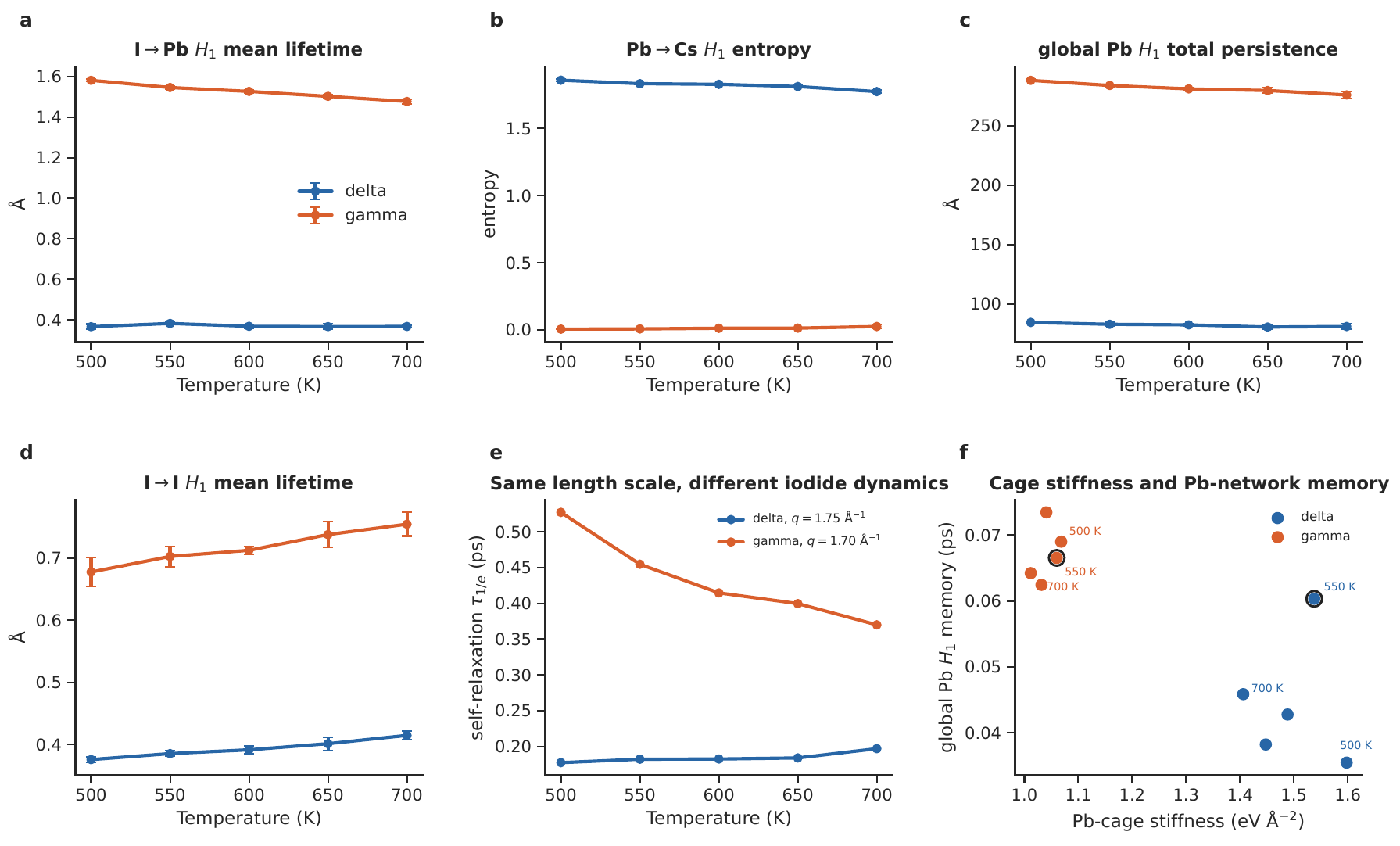}
\caption{\label{fig:bulk}Network topology and dynamics across five temperatures. (a) I$\rightarrow$Pb $H_1$ mean lifetime, (b) Pb$\rightarrow$Cs $H_1$ persistence entropy, and (c) complete-sublattice Pb $H_1$ total persistence distinguish the corner-sharing and chain-like frameworks. (d) I$\rightarrow$I loop lifetime increases on heating in $\gamma$; the accompanying 4.7\% reduction in loop count is reported in the text and SI Figure S5. Points and error bars in a--d are the mean and standard deviation, respectively, of five estimates formed from successive portions of the 64 ps production trajectory. (e) Iodide self-relaxation time, defined by the first $1/e$ decay of $F_s(q,t)$ at each phase's principal iodide-structure-factor peak. (f) Pb-cage stiffness from the force--displacement slope versus the integrated autocorrelation time of complete-sublattice Pb $H_1$; outlined markers denote 550 K.}
\end{figure}

The directed channels distinguish the Pb--I framework from the surrounding Cs cages (Figure~\ref{fig:bulk}a--c). I$\rightarrow$Pb loops persist for 1.48--1.58 \AA\ in $\gamma$ but only 0.37--0.38 \AA\ in $\delta$, reflecting the extended corner-sharing Pb network around iodide in $\gamma$. Pb$\rightarrow$Cs shows the complementary cage structure. Its persistence entropy measures how evenly the total loop lifetime is distributed: the value is large when several loops contribute comparably and approaches zero when one loop dominates. The entropy is 1.78--1.86 in $\delta$ and 0.01--0.03 in $\gamma$, so the Cs environment around Pb contains several comparably persistent loops in the chain phase but one dominant contribution in the perovskite phase. For both quantities, the phase ranges remain separated in every one of the 25 matched comparisons formed from five temperatures and five successive trajectory portions. Chemical directionality therefore separates the topology of the extended Pb framework from that of the Cs cage instead of combining both environments in one undirected distribution.

A diagram built from the complete periodic Pb sublattice provides a center-independent view. Summing the lifetimes of all global Pb loops gives 276--288 \AA\ in $\gamma$ and 80--84 \AA\ in $\delta$ (Figure~\ref{fig:bulk}c). The separation persists across all five temperatures and all successive trajectory segments. Global Pb loop persistence can therefore be used as a compact coordinate for the medium-range connectivity that must be dismantled and rebuilt during the $\delta\leftrightarrow\gamma$ transformation.

Heating changes the topology within the $\gamma$ phase without erasing its separation from $\delta$. From 500 to 700 K, the mean lifetime of I$\rightarrow$I loops increases from 0.68 to 0.75 \AA, while their number decreases by 4.7\% (Figure~\ref{fig:bulk}d). For each prespecified temperature trend, the measured rank correlation was compared with all $5!=120$ possible assignments of the five temperature labels; the permutation probability is the fraction producing a correlation at least as large in magnitude. These probabilities were corrected together by the Benjamini--Hochberg procedure, which limits the expected fraction of false discoveries among the trends called significant. The coupled increase in lifetime and decrease in count has $p_{\mathrm{FDR}}=0.049$. The iodide network therefore does not simply accumulate more cycles on heating: fewer cycles remain, but each survives over a larger interval of connection distance.

The distinction between local motion and network memory emerges from the force and time-correlation data. The primary iodide peaks of the static structure factor occur at $q=1.75$ \AA$^{-1}$ in $\delta$ and 1.70 \AA$^{-1}$ in $\gamma$, corresponding to comparable distances of $2\pi/q=3.59$ and 3.70 \AA. At this length scale, the self-intermediate scattering function measures the fraction of an iodide's initial positional correlation retained after time $t$. Its $1/e$ decay is 2.27 times slower in $\gamma$ than in $\delta$ when the five temperatures are compared pairwise (Figure~\ref{fig:bulk}e and SI Figure S15). In contrast, the force opposing Pb--I displacement gives a larger local stiffness in $\delta$ (1.41--1.60 eV \AA$^{-2}$) than in $\gamma$ (1.01--1.07 eV \AA$^{-2}$). The softer $\gamma$ cage permits larger and more heterogeneous iodide excursions, but the autocorrelation of the complete Pb-loop network decays 1.63 times more slowly in the median temperature-matched comparison (Figure~\ref{fig:bulk}f). Thus displacement amplitude, local restoring force, and loss of medium-range organization are different physical quantities: the corner-sharing network accommodates larger cage motion while retaining its positional and topological organization for longer.

At 550 K, the $\delta$ Pb-loop memory rises to 0.060 ps from 0.036--0.046 ps at the other four temperatures and approaches the $\gamma$ value of 0.067 ps. Estimates from successive trajectory portions still span a common range between the phases, so the observation is a temperature-localized change in network dynamics rather than a separate thermodynamic transition measurement. At the same temperature, the two geometry-conditioned phase contrasts described below reach their minima. This concurrence defines the 550 K anomaly in terms of Pb-network organization within the known $\delta$--$\gamma$ competition range.

\subsection{Network topology retains phase information beyond the tested crystallographic basis}

Topology and conventional crystallographic coordinates describe the same atomic structure from different viewpoints. To establish where they agree and where they differ, we constructed a reference basis using Pyrovskite,\cite{Stanton2023Pyrovskite} PDynA,\cite{Liang2023PDynA} and independently calculated cell and coordination measures. Twenty-seven variables describe the simulation cell, coordination numbers, Pb--I and Cs--I distances, Pb--I--Pb bridges, framework connectivity, and cation off-centering. Ten properties are then evaluated for each of the 64 PbI$_6$ octahedra and represented by their mean, standard deviation, minimum, lower quartile, median, upper quartile, and maximum. These $10\times7=70$ distribution values give 97 variables in total; every definition and unit is listed in SI Tables S5--S6.

\begin{figure}[!htbp]
\centering
\includegraphics[width=\textwidth]{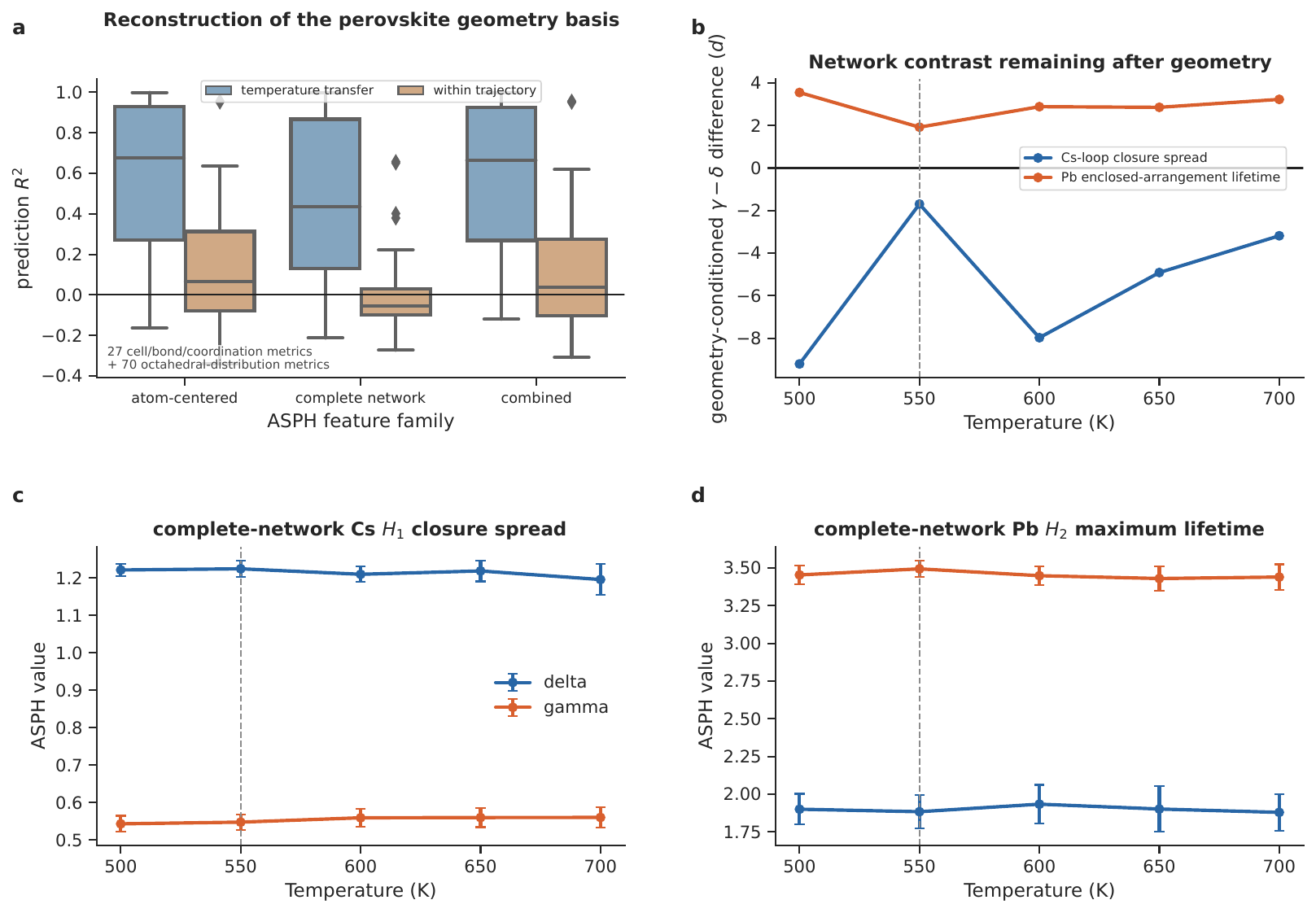}
\caption{\label{fig:geometry}Geometry encoded by, and complementary to, topology. (a) Prediction of the 97 crystallographic quantities when one complete temperature is excluded from training. Boxes summarize scores for 27 cell, bond, coordination, and connectivity quantities and seven distribution statistics for each of ten octahedral measures. The within-trajectory control tests the much smaller fluctuations inside a fixed phase and temperature. (b) The $\gamma-\delta$ difference remaining after all 97 geometric quantities have been used to predict each topological feature, divided by the pooled standard deviation of the five trajectory-segment estimates; zero would mean no remaining phase contrast. Both selected complete-sublattice features retain the same sign at every temperature, with their smallest geometry-conditioned separation at 550 K. Their unadjusted profiles are shown for (c) the standard deviation of Cs $H_1$ death scales and (d) the maximum Pb $H_2$ lifetime. Points and error bars in c and d are means and standard deviations over the five successive trajectory-segment estimates.}
\end{figure}

Local ASPH recovers much of this crystallographic basis at temperatures excluded from fitting, with a median coefficient of determination $R^2=0.68$ across the 97 quantities (Figure~\ref{fig:geometry}a and SI Figures S6--S10). Here $R^2$ is the fraction of the variation at the excluded temperature reproduced by the prediction. A separate ridge regression predicts each crystallographic quantity from the local ASPH features. In each of five validation cycles, these regressions are fitted using both phases at four temperatures and applied to the excluded fifth temperature. This leave-one-temperature-out design tests transfer across thermal conditions rather than interpolation between neighboring frames. The mapping is physically recognizable: I$\rightarrow$Cs connectivity follows the Cs--I distance ($\rho=0.77$), I$\rightarrow$Pb connectivity follows the short Pb--I distance ($\rho=0.72$), and Cs$\rightarrow$I cavity persistence decreases as octahedral angular disorder increases ($|\rho|=0.61$--0.63; SI Figure S9). Spearman's $\rho$ measures monotonic association, with magnitude one denoting perfect rank ordering and zero denoting no rank association. Persistent homology therefore recovers much of the conventional crystallographic description without requiring a PbI$_6$ unit to be assigned before the calculation.

The same ASPH-to-crystallography regressions reproduce little of the small frame-to-frame variation within a fixed phase and temperature (median $R^2=0.07$). The correspondence is therefore specific: ASPH carries the robust structural variation associated with phase and thermal state, not every instantaneous vibration of an individual bond or angle.

To test whether topology retains information not reproduced by these crystallographic quantities, we reverse the prediction direction. In each leave-one-temperature-out cycle, a separate ridge regression predicts each topological feature from the 97 measured variables. Ridge regression penalizes large coefficients, which stabilizes the fit when input variables are correlated. The prediction for the excluded temperature is then subtracted from the observed topological value. Two complete-sublattice quantities retain the same phase difference after this subtraction at all five temperatures and in all 25 comparisons between corresponding trajectory segments (Figure~\ref{fig:geometry}b--d). The standard deviation of Cs $H_1$ death scales is larger in $\delta$ and measures the heterogeneous distances at which loops close around the chain-separated Cs sublattice. The maximum lifetime of complete-sublattice Pb $H_2$ is larger in $\gamma$ and measures the most persistent enclosed arrangement formed by the three-dimensional Pb network. A topological ``cavity'' here is a connected enclosed feature that exists over a finite interval of the filtration; it need not coincide with a physical pore. These phase differences quantify medium-range network organization that is not reproduced by ridge regression from the 97 crystallographic variables.

Both geometry-conditioned phase differences are smallest at 550 K, at the lower edge of the 548--599 K interval measured by calorimetry.\cite{Dastidar2017} The simultaneous maximum in $\delta$ Pb-network memory links this reduced topological separation to slower reorganization of the chain-phase network. The temperature dependence therefore identifies a specific network-level signature near the measured transition temperature: the two phases remain structurally distinct, while the $\delta$ Pb network becomes dynamically more persistent and two medium-range contrasts narrow.

\subsection{Rare network environments precede enhanced Pb motion}

Retaining a separate chemically directed persistence diagram for each Pb center and sampled time converts the network measure into a site-resolved trajectory coordinate without changing its chemical direction or filtration-length interpretation.

\begin{figure}[!htbp]
\centering
\includegraphics[width=\textwidth]{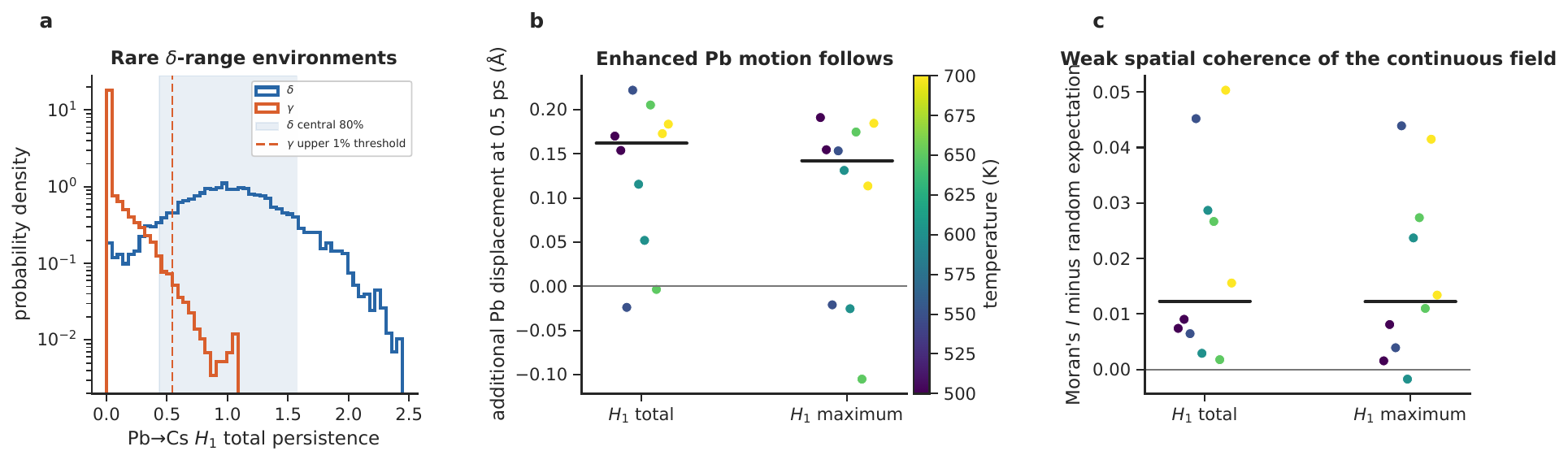}
\caption{\label{fig:precursor}Site-resolved topological susceptibility. (a) Distribution of Pb$\rightarrow$Cs $H_1$ total persistence at 600 K. The blue band is the central 80\% of the $\delta$ values and the dashed orange line is the upper 1\% threshold of the $\gamma$ values; both are fixed using the first three separated 1 ps windows. Events are $\gamma$ observations above the dashed line that also fall inside the blue band. (b) Difference in 0.5 ps Pb displacement associated with the event after adjustment for instantaneous force, speed, local geometry, atom identity, and frame; positive values denote enhanced motion. (c) Moran's spatial autocorrelation of the continuous Pb$\rightarrow$Cs persistence field after subtraction of the random-placement expectation; zero denotes no excess nearest-neighbor correlation. In b and c, color gives temperature, each point is one later window, and the black horizontal segment is the median over all ten windows. Only the first 0.5 ps of the final 1 ps window supports the future-displacement response (SI Section S5).}
\end{figure}

Approximately 1\% of the Pb environments sampled in $\gamma$-\cspbi\ have Pb$\rightarrow$Cs loop persistence within the central 80\% of the $\delta$ distribution (Figure~\ref{fig:precursor}a). This overlap means that, for this specific network coordinate, the Cs cage around a $\gamma$-phase Pb temporarily adopts the loop-persistence scale characteristic of the chain phase. The event describes a local cage fluctuation, not a bulk phase label: it identifies an unusual way in which the surrounding Cs sites connect as the filtration distance grows.

The site analysis uses an 8.0 \AA\ neighborhood because it contains the central Pb and exactly the eight Cs sites that form its immediate A-site cage in all but four of the 160,000 sampled $\gamma$-phase environments; it does not extend into the next Cs shell. Every Pb atom is evaluated in five separated 1 ps windows across each 64 ps trajectory. Data from the first three windows determine which topological quantities distinguish the phases consistently across all five temperatures and fix the numerical threshold defining the rare $\gamma$ tail. Those definitions are then applied unchanged to the final two windows. The result is ten later-window comparisons---two at each of five temperatures---in which event definition and response evaluation use different portions of the trajectories.

The local network fluctuation precedes enhanced motion of the central Pb. For each atom, its displacement over the 0.5 ps after an event is compared with its displacement after ordinary environments, so a persistently mobile site is compared with itself. Events defined by the summed Pb$\rightarrow$Cs loop lifetime precede a median 0.17 \AA\ increase in Pb displacement. The increase has the same positive direction in all ten later-window comparisons and $p_{\mathrm{FDR}}=0.038$ after correction across the tested topology--response pairs. The maximum individual loop lifetime gives a 0.13 \AA\ increase with the same ten-of-ten direction. A second analysis simultaneously includes force, speed, Pb off-centering, Pb--I bond distortion, cage size, atom identity, and frame identity. The remaining median displacement increases are 0.16 and 0.14 \AA, positive in eight and seven comparisons, respectively (Figure~\ref{fig:precursor}b; SI Section S5, Figure S11, and Tables S8--S9). The rare-event support ranges from one to eight and two to nine distinct Pb sites per comparison for the two quantities, and is listed explicitly in SI Table S9.

The unusual cage values remain localized rather than forming extended regions. Moran's $I$ compares the similarity of the continuous Pb$\rightarrow$Cs persistence values on neighboring Pb sites with the similarity obtained after randomly reassigning those values over the same Pb-neighbor graph. The median excess above random placement is 0.012 (Figure~\ref{fig:precursor}c), indicating only weak nearest-neighbor smoothness. The physical result is therefore site resolved: a transient change in Cs-cage connectivity marks Pb atoms that move farther during the following 0.5 ps, before any trajectory-wide structural change is invoked.

\subsection{A network coordinate resolves a structurally mixed phase boundary}

Evaluating Pb-loop persistence at every Pb site produces a spatial field whose microscopic definition is unchanged across the two bulk regions and the intervening boundary. Spatial averaging is applied only within narrow slabs when the field is plotted.

A spatial description of a phase boundary should connect the characteristic values of the two bulk phases and resolve the intervening mixed structure. We constructed a periodic $\delta|\gamma$ cell from representative 600 K AIMD configurations. The two complete 320-atom parent cells were repeated parallel to the interface and strained symmetrically to a common two-dimensional lattice. The relative lateral translation of the two slabs is called the registry. Nine registries were compared using the same relaxation procedure, and the relaxed structure with the lowest energy was used for the trajectory. The resulting 2560-atom cell contains two periodic interfaces, each with an area of approximately 1366 \AA$^2$; the $\delta$ and $\gamma$ slabs are 43.28 and 50.49 \AA\ thick, respectively, and the maximum in-plane strain is 3.3\% (SI Section S6 and Figure S12).

The 2560-atom cell retains bulk-like regions farther than the descriptor range from both periodic interfaces. Its relaxation and finite-temperature dynamics were calculated with a MACE-MH-0 machine-learned interatomic potential initialized from an energy model pretrained on MatPES r2SCAN calculations.\cite{Batatia2022MACE,Batatia2025CrossLearning} Model weights were fitted to 1110 structures spanning all ten bulk AIMD conditions. A separate set of 400 structures was used to choose the learning rate and the relative weights of energy, force, and stress errors during fitting. Those choices were then fixed, and errors were evaluated once on a third set of 400 structures drawn from different portions of the trajectories. On this final set, the potential gives a relative-energy mean absolute error of 8.9 meV atom$^{-1}$, a force root-mean-square error of 0.014 eV \AA$^{-1}$, a stress root-mean-square error of 0.12 GPa, and a $\delta$--$\gamma$ energy-difference error of 0.6 meV atom$^{-1}$. The relaxed interface was propagated for 20 ps at 600 K (SI Table S10).

\begin{figure}[!htbp]
\centering
\includegraphics[width=\textwidth]{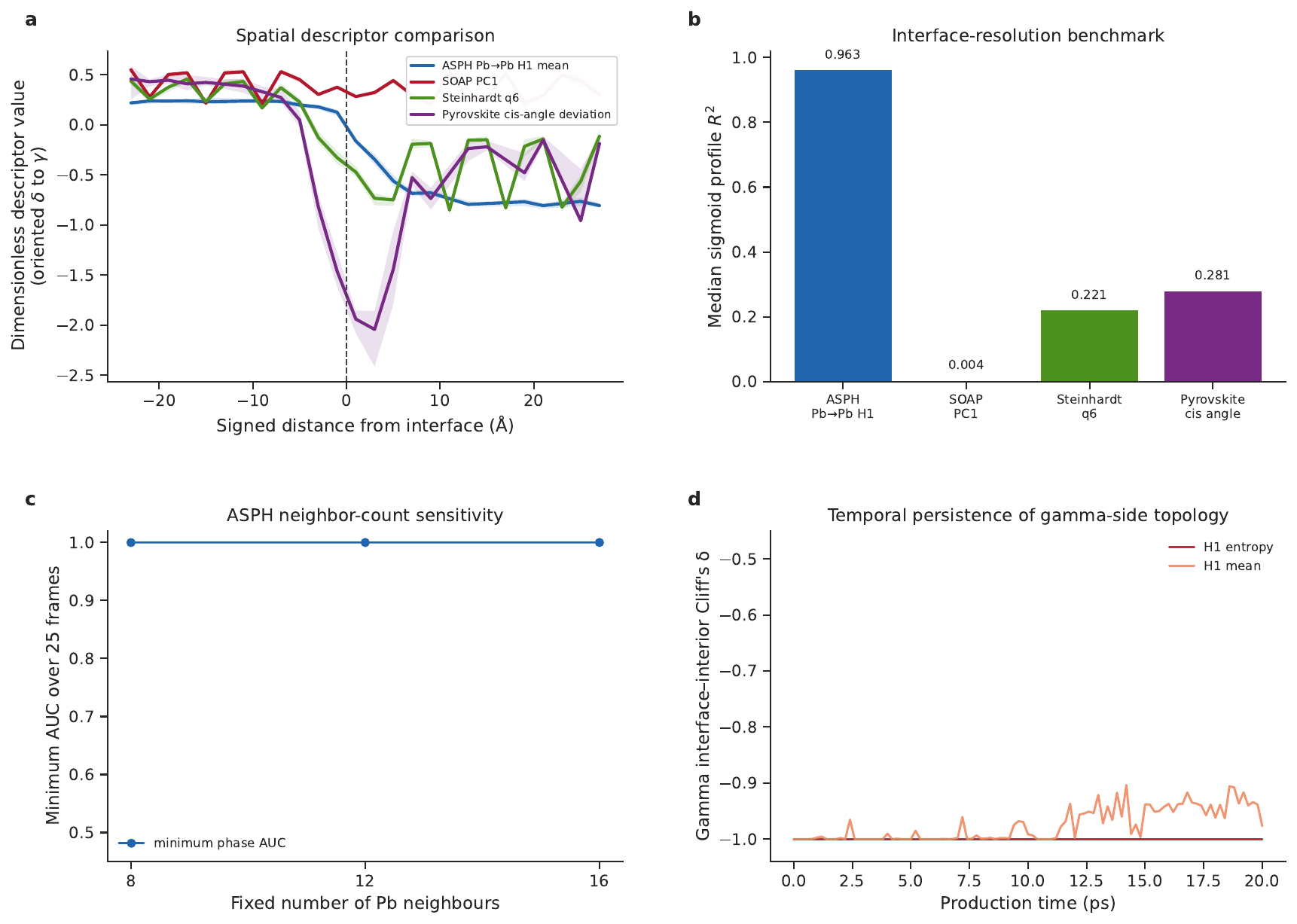}
\caption{\label{fig:interface}A continuous structural field across a 2560-atom $\delta|\gamma$ interface. (a) Signed distance is negative on the constructed $\delta$ side and positive on the $\gamma$ side; zero marks the interface plane. Within each frame, every field is oriented from its $\delta$ to $\gamma$ interior value and divided by its interquartile range, the span containing the central 50\% of values. Lines and bands are temporal medians and central-50\% ranges in 2 \AA\ slabs. (b) Median coefficient of determination from fitting each site-resolved scalar field in 25 matched frames to one sigmoidal boundary profile; unity is a perfectly resolved monotonic profile. SOAP PC1 is the first unsupervised principal component, or direction of greatest variance, fitted separately to the SOAP vectors in each frame. (c) Minimum bulk-phase area under the receiver-operating-characteristic curve over 25 frames for Pb$\rightarrow$Pb $H_1$ neighborhoods containing 8, 12, or 16 Pb neighbors; the area is the probability of correctly ordering one randomly chosen site from each bulk region. (d) Difference between the $\gamma$-side interface and interior distributions over 101 frames, measured by Cliff's $\delta$; negative values denote reduced loop persistence at the boundary.}
\end{figure}

All nine directed Cs/Pb/I pairs and their $H_0$--$H_2$ summaries were evaluated across the interface. Using the same fixed-16 $H_1$ lifetime mean for every pair, all nine fields distinguish the two bulk-like regions in every one of the 25 comparison frames; their worst-frame phase AUCs range from 0.754 to 1.000 (SI Table S11). Pb$\rightarrow$Pb is used as the primary field because it gives exact phase ordering, has the largest normalized boundary-to-interior change among these nine channels, and directly follows the Pb framework whose connectivity changes between the chain and corner-sharing phases. The Pb-loop field approaches distinct bulk values on the two sides and changes smoothly across the mixed region, giving a median sigmoid-profile $R^2=0.963$ and a minimum of 0.953 over the trajectory (Figure~\ref{fig:interface}a,b). Here $R^2=1$ denotes a scalar field described exactly by one smooth transition between its two bulk values. The result is unchanged for neighborhoods containing 8, 12, or 16 Pb atoms (Figure~\ref{fig:interface}c and SI Figure S12).

The same frames were analyzed with crystallographic, bond-orientational, and density-based quantities. Pyrovskite supplies ten PbI$_6$ measures: Pb--I bond mean, spread, and range; bond-length and bond-angle distortion; three off-centering measures; and cis- and trans-angle deviations.\cite{Stanton2023Pyrovskite} The best of these is the cis-angle deviation, with median profile $R^2=0.281$. Steinhardt $q_6$, a measure of local bond-orientational order, gives 0.221. Standard SOAP produces a 1197-component vector describing the Cs, Pb, and I neighbor densities at each site. To obtain one spatial value per site without using the phase labels, principal-component analysis identifies the linear combination with the greatest variance separately in each frame. This first SOAP component gives $R^2=0.004$: its dominant variation is not aligned with distance across the boundary. In contrast, the chemically defined Pb-loop lifetime is already a scalar and follows the change in Pb-network connectivity. Figure~\ref{fig:interface}a shows all four quantities averaged in 2 \AA\ slabs parallel to the boundary and plotted against distance normal to it.

The boundary also reveals why a network coordinate is useful when a local reference motif becomes incomplete. PDynA calculates an octahedral mode only after matching six iodides to the sites of a reference PbI$_6$ octahedron. The Pb-layer density normal to the boundary provides the spatial definitions. The interval within 4 \AA\ contains exactly the boundary-facing $\gamma$ Pb layer at both interfaces in every analyzed frame; the next $\gamma$ Pb layer begins at least 4.6 \AA\ from the boundary. The same normal-distance interval is applied on the $\delta$ side. The interior comparison begins at 9.6 \AA, after the three Pb layers closest to each boundary have been excluded; results retain their direction when this interior margin is changed to 8.0 or 11.2 \AA\ (SI Section S6). Octahedral matching succeeds for only 40\% of boundary-adjacent Pb sites originating from $\delta$ and 7\% of those originating from $\gamma$, whereas the corresponding interior fractions are 100\% and 84\%. Thus, most Pb sites nearest the boundary cannot be represented by an ideal-octahedron mode because their six iodides cannot be matched to the reference motif. The Pb-loop coordinate instead follows the Pb network and remains defined at every site.

On the $\gamma$ side, the interfacial Pb-loop lifetime is lower than its bulk-like interior value throughout the 20 ps trajectory. Figure~\ref{fig:interface}d reports Cliff's $\delta$, the probability that a randomly chosen interfacial value exceeds an interior value minus the probability of the reverse ordering. Its temporal median is $-0.993$; a value of $-1$ would mean that every sampled interfacial value is smaller. The nearly complete negative ordering shows that the corner-sharing Pb network loses medium-range loop persistence at the boundary. This supplies the physical content of the spatial profile: the field does not merely locate the interface, but measures how strongly the extended $\gamma$ network is disrupted there.

\subsection{Configurational ordering exposes errors hidden by aggregate energy accuracy}

The same \cspbi\ chemistry provides a fixed-composition test of configurational ordering. Eremin et al. enumerated Cd- and Zn-substituted Pb sublattices in the $\delta$ and $\gamma$ phases using 160-atom cells.\cite{Eremin2024Doped} Their database contains 202 structures with DFT-relaxed formation energies. A 142-structure reference set, containing zero to three substitutions, is used to choose the descriptors, principal-component reduction, and regression penalty. A separate 60-structure test set contains three to five substitutions, including four- and five-dopant cells not represented in the reference set. The database also contains 73,760 additional configurations with SchNet and Allegro energies but no DFT energies. These configurations are excluded from accuracy assessment and are used only after the DFT tests to determine whether each neural-network energy landscape thermally favors compact or dispersed dopant arrangements. At a fixed phase, dopant identity, and substitution count, every structure has the same composition; only the arrangement of substituted sites differs. Their relative energies, rather than the mean energy of the composition, determine which arrangements receive the largest equilibrium populations.

The substituted sites define a direct coordinate for this arrangement. At zero connection distance, each dopant is an isolated group. As the distance increases, the two nearest groups merge, followed by progressively more distant groups, until all dopants belong to one connected network. The finite $H_0$ death scales are the distances at which these mergers occur. Compact dopants merge at short distances and therefore have small $H_0$ scales; dispersed dopants remain separate until larger distances and have large scales. This ordered set of merger distances measures clustering over the complete periodic cell and distinguishes arrangements that share the same composition. Its 28 distribution summaries are defined in SI Section S8.

\begin{figure}[!htbp]
\centering
\includegraphics[width=\textwidth]{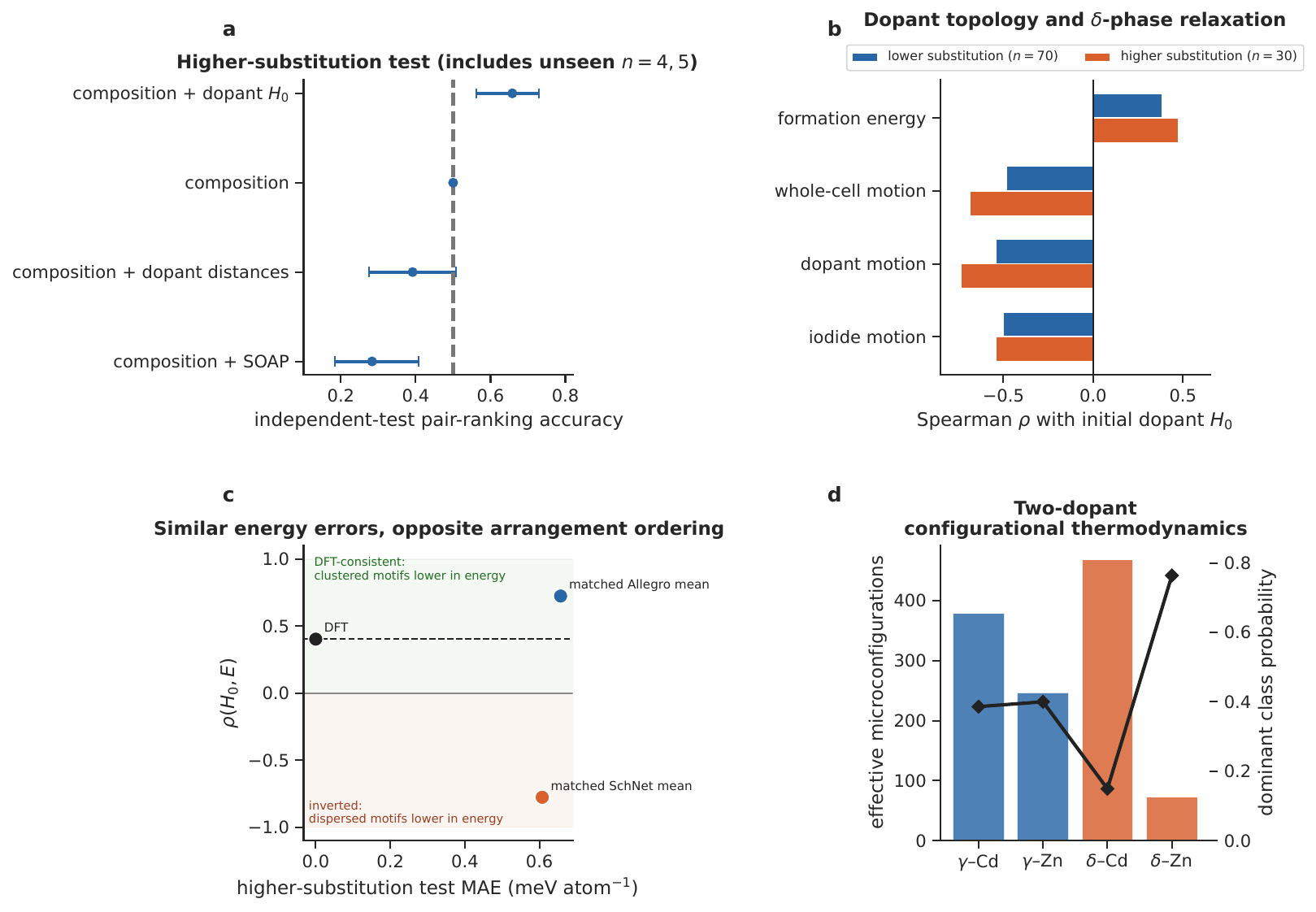}
\caption{\label{fig:doping}Dopant connectivity and configurational physics. (a) Fraction of correctly ordered structure pairs among the 60 higher-substitution DFT structures, including four- and five-dopant cells not represented in the 142-structure reference set. Each regression includes the same phase, dopant identity, and substitution-count terms. Error bars are 95\% intervals obtained by resampling groups with the same phase, dopant species, and substitution count while retaining all structures and pairwise comparisons within each selected group; the dashed line marks the 0.5 score expected from random pair ordering. (b) Spearman correlation between the initial mean dopant $H_0$ connection scale and four DFT relaxation responses in $\delta$. Blue denotes the 70 lower-substitution structures used to identify the relationships and orange the separate 30 higher-substitution structures used to evaluate them; positive energy correlation means dispersed dopants remain higher in energy, whereas negative displacement correlations mean compact dopants relax farther. (c) Correlation between $H_0$ and energy after subtracting the separate mean of each fixed-composition set, plotted against energy mean absolute error on the higher-substitution structures. SchNet and Allegro points are means over 48 fitted models in the matched non-pretrained families trained across both phases and dopants. DFT and Allegro place clustered arrangements lower in energy; SchNet reverses that ordering despite a similar error. (d) Effective number of populated microscopic two-dopant arrangements (bars) and probability of the most populated symmetry-inequivalent class (diamonds) at 300 K, calculated from static DFT energies and symmetry degeneracies.}
\end{figure}

Before examining learned potentials, the DFT structures establish the physical meaning of this coordinate. In the 60-structure test set, larger mean $H_0$---greater dopant separation---is associated with higher final formation energy ($\rho=0.47$). The same dispersed arrangements undergo less whole-cell, dopant, and iodide displacement during DFT relaxation, with $\rho=-0.69$, $-0.74$, and $-0.54$, respectively (Figure~\ref{fig:doping}b). Compact dopants are therefore stabilized together with a larger cooperative relaxation of the surrounding iodide framework. At the atomic scale, an iodide with a larger initial I$\rightarrow$Pb $H_0$ connection scale subsequently moves farther. After the nearest-dopant distance, neighborhood size, and cloud radius are removed from both quantities, the within-structure rank correlation remains 0.38 and is positive in all 30 test-set $\delta$ structures (SI Figure S16). Dopant clustering and the response of individual iodides consequently describe the collective and local parts of the same host relaxation.

A ridge regression tests whether the dopant connectivity also transfers quantitatively. Seven composition terms specify phase, dopant identity, substitution count, its square, and their pairwise interactions; 28 $H_0$ summaries specify arrangement. After standardization on the 142 lower-substitution structures, principal-component analysis reduces these 35 correlated inputs to the five linear combinations retaining 95\% of their variation, and ridge regression maps them to formation energy. Applied unchanged to the 60 higher-substitution structures, it gives $R^2=0.925$ and a mean absolute error of 1.71 meV atom$^{-1}$. More importantly for fixed-composition selection, it orders 66\% of structure pairs correctly. A random ordering, or a model using composition alone and therefore assigning the same energy to every arrangement in a fixed-composition set, scores 50\%; explicit dopant-distance summaries score 39\% and standard SOAP scores 28\% (Figure~\ref{fig:doping}a). The 95\% interval for the $H_0$ advantage over SOAP is 0.16--0.54 when the chemical groups defined by phase, dopant species, and substitution count are resampled while retaining all structures and shared pairs within each group. The regression provides a quantitative test that the physically defined connectivity coordinate transfers to four- and five-dopant cells.

The DFT relation between connectivity and energy provides a separate test of the supplied graph-neural-network predictions. Within every fixed-composition set, we subtract that set's mean energy and mean $H_0$ before calculating Spearman's $\rho$. This removes differences caused by phase, dopant identity, or substitution count and retains only whether compact and dispersed arrangements are ordered consistently. DFT gives $\rho=+0.40$ across all 202 structures and $+0.49$ in the 60-structure test set: within a common composition, dispersed arrangements lie higher in energy.

Eremin et al. supplied eight model families formed from two architectures, two training scopes, and, for Allegro, three initialization choices.\cite{Eremin2024Doped} A ``both--both'' model is trained jointly on Cd and Zn structures in both phases, whereas an ``element--both'' model uses one dopant species in both phases. ``Non-pretrained'' denotes fitting from that \cspbi\ dataset without prior initialization from the Open Catalyst Project or AFLOW. Each supplied prediction table contains 48 fitted models. Figure~\ref{fig:doping}c compares the non-pretrained, both--both SchNet and Allegro families; their training data and initialization are matched, leaving network architecture as the differing model-design choice. The mean Allegro prediction preserves the DFT connectivity--energy ordering ($\rho=+0.72$), while the mean SchNet prediction reverses it ($\rho=-0.78$). Their mean absolute errors on the separate test structures are nevertheless similar: 0.66 and 0.61 meV atom$^{-1}$, respectively. Comparable aggregate errors therefore coexist with opposite ordering along a physically defined configurational coordinate.

The full collection contains 384 models: 288 Allegro and 96 SchNet fits. Among the 247 models with errors below 1 meV atom$^{-1}$ on the separate test structures, 61 reverse the DFT ordering; these comprise 42 of 215 low-error Allegro models and 19 of 32 low-error SchNet models. The 73,760 GNN-only configurations then extend this sign test over the enumerated landscapes. Two phases, two dopants, five substitution counts, and three temperatures define 60 thermodynamic conditions. Across the six Allegro family means, these give 360 family--condition results, of which 300 favor compact arrangements. Both SchNet family means favor more dispersed arrangements in all 120 corresponding results (SI Section S8). An energy model can therefore achieve a small average error while favoring the wrong class of arrangements at fixed composition. The topology--energy relation supplies a fixed-composition fidelity test for both material-specific and universal machine-learned potentials, whose aggregate errors need not reveal reversal of configurational ordering. This relation also provides a basis for structure-aware sampling, validation, and training of machine-learned potentials.

The complete two-dopant DFT spaces show the thermodynamic consequence of this ordering. Each phase--dopant case contains 15 or 16 symmetry-inequivalent classes representing 496 microscopic arrangements once symmetry multiplicities are counted. Following the 300 K thermal scale used by Eremin et al. for configurational averaging,\cite{Eremin2024Doped} we assign each class the canonical probability $p_i\propto g_i\exp[-(E_i-E_{\min})/(k_BT)]$, where $g_i$ is its number of symmetry-equivalent realizations. The configurational entropy gives an effective number of populated microscopic arrangements, $N_{\rm eff}=\exp(S_{\rm conf}/k_{\rm B})$. If all arrangements had equal energy, their populations would follow only their symmetry multiplicities. Including the DFT energy differences shifts the mean $H_0$ below this equal-energy value by 0.40 \AA\ for $\delta$--Cd and 4.30 \AA\ for $\delta$--Zn; the negative shift means that thermal population moves toward more compact dopant networks.

The magnitude of this selection is chemically distinct. At 300 K, $\delta$--Cd retains $N_{\rm eff}=468$ of the 496 available microscopic arrangements, whereas $\delta$--Zn retains only 73 and one Zn class carries 76\% of the total probability (Figure~\ref{fig:doping}d). Static DFT energetics therefore predict broad Cd disorder but strong selection of a compact Zn arrangement, consistent with the larger cooperative relaxation associated with clustered dopants. Reversing the $H_0$--energy relation reverses this population shift: Boltzmann weighting would favor dispersed structures where DFT favors clustered ones, changing the dominant configurations and their configurational entropy. The topology--energy sign is consequently a thermodynamic fidelity test, not merely another correlation metric.

\section{Conclusions}

Chemical network topology separates the amplitude of local motion from the persistence of medium-range organization. In $\gamma$-\cspbi, the softer Pb--I cage permits larger and more heterogeneous iodide excursions than in $\delta$, while iodide positional correlations decay 2.27 times more slowly and the Pb network retains topological memory 1.63 times longer. The corner-sharing network therefore accommodates substantial local motion without losing its collective organization at the same rate. At 550 K, the simultaneous increase in $\delta$ Pb-network memory and narrowing of two geometry-conditioned phase contrasts identifies a network-level anomaly within the known phase-competition range.

The same connectivity scale remains informative below the bulk average and across a structurally mixed boundary. Rare $\gamma$-phase Pb environments that enter the $\delta$ range of Cs-cage loop persistence precede 0.17 \AA\ greater Pb displacement over the next 0.5 ps. Across the 2560-atom $\delta|\gamma$ cell, Pb-loop lifetime remains defined where an ideal PbI$_6$ assignment becomes incomplete and shows that the corner-sharing network loses medium-range persistence at the boundary. These results connect a named structural fluctuation to later local motion and to the spatial disruption of the same chemical network.

In substituted \cspbi, compact dopant networks undergo greater cooperative iodide relaxation and lie lower in DFT energy than dispersed arrangements of the same composition. This ordering produces broad configurational disorder for two Cd substitutions but concentrates 76\% of the $\delta$--Zn population in one symmetry-inequivalent class at 300 K. SchNet and Allegro families with comparable mean energy errors encode opposite connectivity--energy relations, and 61 of 247 supplied models below 1 meV atom$^{-1}$ error reverse the DFT sign. Because same-composition energy differences determine Boltzmann populations, aggregate error alone does not establish configurational fidelity. Across all four resolutions, each reported topological quantity retains its chemical direction, homology dimension, and filtration length, so its physical meaning does not have to be inferred from a fitted projection. A chemically interpretable network coordinate complements energy and force benchmarks by testing whether a learned potential preserves the structural ordering that selects the thermodynamic ensemble.

\section{Methods}

\subsection{AIMD trajectories}

The $\delta$ and $\gamma$ simulation cells each contained Cs$_{64}$Pb$_{64}$I$_{192}$. Separate trajectories were calculated for both phases at 500, 550, 600, 650, and 700 K. AIMD used the CP2K Quickstep Gaussian-and-plane-wave implementation,\cite{Kuhne2020CP2K} the Perdew--Burke--Ernzerhof (PBE) exchange--correlation functional,\cite{Perdew1996PBE} Grimme D3 dispersion with Becke--Johnson damping,\cite{Grimme2010D3,Grimme2011BJ} Goedecker--Teter--Hutter pseudopotentials, triple-$\zeta$ valence plus polarization short-range molecularly optimized basis sets (TZVP-MOLOPT-SR), a 400 Ry auxiliary density cutoff, and five multigrid levels. Fully flexible constant-number, constant-pressure, constant-temperature (NPT) dynamics used a 2 fs integration step, a global Nos\'e--Hoover chain thermostat with a 100 fs time constant, and a barostat at 0.1 MPa with a 500 fs time constant. Coordinates, cells, forces, velocities, energies, and analytic stress were recorded. Convergence of temperature, energy, volume, and pressure after excluding 5, 10, 15, or 20 ps supports a common 10 ps equilibration period: every 10 ps estimate differs from its 20 ps counterpart by less than 0.55 times their pooled uncertainty estimated from autocorrelation-corrected 0.5 ps time blocks. SI Sections S1--S2 give the trajectory identities, equilibration checks, and sampling schedules used for the bulk frame comparison, atom-resolved analysis, and time-correlation functions.

\subsection{Persistent-homology construction}

For local $A\rightarrow B$ \ASPH, the minimum-image point cloud $X_i^{A\rightarrow B}(R)$ contains center $i$ of species $A$ at the origin and every atom of species $B$ within the primary radius $R=9.5$ \AA. This radius lies immediately below the 9.527 \AA\ minimum half-translation of the analyzed periodic cells. Vietoris--Rips complexes $\mathrm{VR}(X,\epsilon)$ were constructed with Ripser.py\cite{Tralie2018Ripser} while the connection distance $\epsilon$ was increased. Every finite topological feature is represented by its birth and death scales, $(b_k,d_k)$, and lifetime $\ell_k=d_k-b_k$. We retain $H_0$, $H_1$, and $H_2$ and summarize each diagram by feature count; moments and extrema of birth, death, and lifetime; total persistence $\sum_k\ell_k$; and persistence entropy $-\sum_k p_k\log p_k$, where $p_k=\ell_k/\sum_j\ell_j$. Complete-network diagrams were calculated from the periodic Cs, Pb, or I sublattice rather than from one atomic center. Matched $R=8.0$ \AA\ calculations test the radial stability of the bulk phase contrasts; the site-resolved Pb calculation uses the same radius to isolate the immediate Pb--Cs cage. Complete definitions are given in SI Section S3 and Tables S1--S3.

\subsection{Geometry, dynamics, and statistical design}

The geometry basis combines Pyrovskite,\cite{Stanton2023Pyrovskite} PDynA,\cite{Liang2023PDynA} and independently calculated cell and coordination quantities. Features with more than 5\% missing assignments or no variation were excluded before regression. To test temperature transfer, a separate ridge regression predicts each target quantity after fitting on four temperatures; the fifth temperature is excluded in turn until each has served once as the test. A separate within-trajectory comparison fits and evaluates the regressions on nonoverlapping temporal portions of the same phase and temperature. Ridge regression is a linear method whose penalty on large squared coefficients limits overfitting; the penalty strength is selected using only the fitting portion of each cycle. To measure topological information not reproduced by the geometry basis, each topological feature is predicted from all geometric quantities at the four fitting temperatures, and this prediction is subtracted from the observed value at the excluded temperature.

Iodide relaxation was measured with the self-intermediate scattering function at the primary peak of the iodide static structure factor. The reported relaxation time is the first decay to $1/e$ after normalization by the long-time plateau. Cage stiffness is the magnitude of the local slope relating Pb--I displacement to the opposing force. Statistical comparisons retain the largest natural sampling unit: a phase--temperature condition, a specified trajectory segment or window, or a complete substituted structure. Atoms and frames are not treated as independent replicates. When several quantities address one question, their test probabilities are adjusted with the Benjamini--Hochberg procedure; $p_{\mathrm{FDR}}$ denotes the adjusted probability. This procedure controls the expected fraction of false positives among the results declared significant. Confidence intervals for trajectory quantities are calculated by resampling contiguous trajectory segments, keeping neighboring frames together. For dopant-ordering accuracy, structures are grouped by phase, dopant species, and substitution count; resampling retains every structure and every pairwise comparison within a selected group. Full equations and validation layouts are provided in SI Sections S3--S5 and S7.

\subsection{Prospective site-resolved analysis}

At each temperature, only the first three separated 1 ps windows are used to find channels with the same phase-difference direction across all five temperatures. For a selected channel, its upper 1\% tail in one phase defines rare events; an event is retained when its value also lies within the central 80\% of the other phase. These choices are fixed before the final two windows are examined. For a Pb event at time $t$, the response is its displacement from $t$ to $t+0.5$ ps after subtracting the mass-weighted mean displacement of all 320 atoms, thereby removing translation of the simulation cell as a whole. The primary effect compares event and ordinary observations for the same atom and then summarizes the contrasts over the ten later-window tests. Sensitivity regressions include a separate baseline for each atom and frame together with instantaneous speed, force magnitude, Pb off-centering, PbI$_6$ bond distortion, number of neighbors, and cloud radius. Spatial coherence of the continuous persistence field is measured by Moran's $I$ on a periodic six-nearest-Pb graph and reported after subtraction of its random-placement expectation. SI Section S5 provides the formula, complete selection procedure, and multiple-test correction.

\subsection{Interface and doped calculations}

The interface was assembled from complete 600 K AIMD cells selected before any descriptor was evaluated. Integer in-plane repeats were searched for a common two-dimensional lattice, the mismatch strain was shared between the phases, and nine lateral alignments were compared using the same relaxation settings. A MACE machine-learned interatomic potential, initialized from the MACE-MH-0 energy head pretrained on MatPES r2SCAN calculations and then fine-tuned on bulk AIMD energies, forces, and stresses, supplied relaxation and fixed-number, fixed-volume, fixed-temperature (NVT) dynamics for the 2560-atom cell. Separate trajectory intervals served three roles: 1110 structures fitted the model weights, 400 selected the learning rate and relative energy--force--stress weights, and a third set of 400 supplied the final errors reported above. The calculated observables are structural profiles along the resulting interface trajectory. At each saved time, atoms were binned by their signed distance normal to the interface, and a sigmoid was fitted independently to each descriptor profile to measure whether it forms one smooth transition between the two bulk values. SOAP used the DScribe implementation\cite{Himanen2020DScribe} with Cs, Pb, and I neighbor densities, $r_{\rm cut}=6$ \AA, and $n_{\max}=l_{\max}=6$.

The doped analysis uses the DFT structures and the SchNet and Allegro predictions supplied by Eremin et al.\cite{Eremin2024Doped} Atom-centered doped-cell topology uses an 8.5 \AA\ radius, the largest one-decimal radius below half of the shortest lattice translation across the 160-atom database. Descriptor choice and regularization use the 142 lower-substitution structures; the fitted procedure is then applied unchanged to the 60 higher-substitution structures. Canonical populations are calculated from the 160-atom DFT cell energies and the supplied symmetry degeneracies. Detailed interface construction and potential validation are provided in SI Section S6; doped-data partitions, ranking definitions, and thermodynamic equations are given in SI Section S8.

\section*{Data and Code Availability}

The data and analysis code supporting this study are available from the corresponding authors upon reasonable request.

\section*{Supporting Information}

The Supporting Information provides additional computational details and validation for the bulk, site-resolved, interface, and doped-structure analyses.

\bibliographystyle{unsrtnat}
\bibliography{references}

\end{document}


\begin{center}
{\LARGE\bfseries Supporting Information}\par
\vspace{6pt}
{\Large Chemically Resolved Topological Coordinates Link Structural Dynamics and Configurational Thermodynamics}\par
\vspace{8pt}
Ayush Kumar Pandey$^{1*}$ and Abhishek Tewari$^{2,3*}$\par
\small $^1$Department of Physics, Indian Institute of Technology Roorkee, Uttarakhand, India\par
$^2$Department of Metallurgical and Materials Engineering, Indian Institute of Technology Roorkee, Uttarakhand, India\par
$^3$Mehta Family School of Data Science and Artificial Intelligence, Indian Institute of Technology Roorkee, Uttarakhand, India\par
$^*$Corresponding authors: \texttt{ayush\_p@ph.iitr.ac.in}; \texttt{abhishek@mt.iitr.ac.in}
\end{center}

\section{Topological constructions and physical observables}

The analyses use one chemically directed topological construction at four resolutions. A diagram for a complete chemical sublattice describes the bulk network; a diagram centered on one atom describes its environment; spatial averages of the atom-centered values describe an interface; and a diagram containing only substitutional sites describes dopant arrangement. The bulk trajectories are first used to determine which topological quantities correspond to familiar crystallographic distortions and which retain organization over several coordination shells. The atom-centered values are then tested against later motion, the spatial averages against a large phase boundary, and the dopant values against density-functional-theory (DFT) energy ordering and relaxation. Forces, velocities, stresses, octahedral modes, and time-correlation functions are used as independent physical measurements rather than as additional structural labels. Atom-specific persistent homology is abbreviated ASPH. The bulk trajectories employ a constant-number, constant-pressure, constant-temperature ensemble with a fully flexible cell (NPT).

\begin{table}[h]
\centering
\caption{\label{tab:traj}AIMD state points and uniform analysis design. Each trajectory contains 10 ps equilibration and 64 ps production.}
\small
\begin{tabular}{cccccc}
\toprule
Phase & $T$ (K) & simulated time (ps) & atoms & ensemble & trajectory estimates\\
\midrule
$\delta$ & 500 & 74 & 320 & flexible-cell NPT & 5\\
$\delta$ & 550 & 74 & 320 & flexible-cell NPT & 5\\
$\delta$ & 600 & 74 & 320 & flexible-cell NPT & 5\\
$\delta$ & 650 & 74 & 320 & flexible-cell NPT & 5\\
$\delta$ & 700 & 74 & 320 & flexible-cell NPT & 5\\
$\gamma$ & 500 & 74 & 320 & flexible-cell NPT & 5\\
$\gamma$ & 550 & 74 & 320 & flexible-cell NPT & 5\\
$\gamma$ & 600 & 74 & 320 & flexible-cell NPT & 5\\
$\gamma$ & 650 & 74 & 320 & flexible-cell NPT & 5\\
$\gamma$ & 700 & 74 & 320 & flexible-cell NPT & 5\\
\bottomrule
\end{tabular}
\normalsize
\end{table}

\section{Trajectory identity and equilibration}

\subsection{Trajectory identity}

Temperature assignments were read from the CP2K \texttt{TEMPERATURE} field, and phase assignments follow the structure referenced by \texttt{COORD\_FILE\_NAME}. All cells have Cs$_{64}$Pb$_{64}$I$_{192}$ stoichiometry. Duplicate records at restart boundaries were removed before time indexing.

\subsection{Equilibration and production windows}

The equilibration interval was chosen by repeating the production averages after discarding the first 5, 10, 15, or 20 ps. Temperature, potential energy, volume, and pressure were examined for all ten phase--temperature conditions. Uncertainty was estimated from 0.5 ps time-block means, with their effective number reduced when successive blocks were autocorrelated. With a 10 ps exclusion, no mean differs from its value after a 20 ps exclusion by more than 0.55 times the square root of the summed squared standard errors of the two estimates. Because the underlying time intervals overlap, this comparison is a convergence diagnostic rather than a test between independent samples. Ten picoseconds also spans 20 barostat time constants. We therefore use 10 ps as a common equilibration interval and retain 64 ps for production.

Three sampling layouts serve different analyses. The systematic bulk comparison uses 25 frames at equal intervals through each production trajectory, one approximately every 2.7 ps and 250 frames across the ten state points. This interval is about 40 times the longest measured integrated memory of the headline complete-network coordinate (0.067 ps). In chronological order, consecutive groups of five sampled frames are averaged to provide one estimate for each successive fifth of the trajectory. The resulting five estimates test whether a phase contrast persists from early to late production. The atom-centered site analysis uses 100 consecutive frames (1 ps) beginning at production times 0, 15.75, 31.50, 47.25, and 63.00 ps. The first three separated windows define the rare-environment rule and its numerical threshold; the final two evaluate later motion using that fixed rule. Time-correlation functions use the continuous production trajectory, and scattering functions are also calculated in four nonoverlapping 16 ps segments. These layouts answer different questions, and atoms or individual frames are never counted as separate trajectory replicates.

Statistical tests preserve these trajectory-level sampling units. For a monotonic temperature trend, the observed Spearman rank coefficient is compared with all $5!=120$ possible orderings of the five temperature values; the resulting probability does not rely on a large-sample approximation. When several topological quantities test the same question, the Benjamini--Hochberg procedure is applied to their individual probabilities. It controls the expected fraction of false positives among the results declared significant; the adjusted probability is denoted $p_{\mathrm{FDR}}$. Spearman's $\rho$ measures monotonic rank association and ranges from $-1$ for exactly opposite ordering to $+1$ for identical ordering. The coefficient of determination $R^2$ is the fraction of the observed variation reproduced by a prediction. Confidence intervals for trajectory quantities are calculated by resampling contiguous trajectory segments, keeping neighboring frames together. For dopant pair ranking, structures are grouped by phase, dopant species, and substitution count. Whenever one such group is selected during resampling, all of its structures and all pairwise comparisons among them are retained. This prevents correlated frames, or different pairs that share the same structure, from being counted as independent observations.

\begin{table}[h]
\centering
\caption{\label{tab:equil}Equilibration and temporal sampling used by each analysis.}
\begin{tabular}{p{0.27\textwidth}p{0.63\textwidth}}
\toprule
Analysis & Sampling\\
\midrule
Simulated interval & 0--74 ps for every state point\\
Equilibration & 0--10 ps excluded from production statistics\\
Production & 10--74 ps, identical for all ten trajectories\\
Systematic bulk extraction & 25 evenly spaced frames; five successive estimates of five frames\\
Site-resolved extraction & Five separated 1 ps windows; 100 frames per window\\
Statistical unit & State point or prespecified trajectory segment/window, never atom or frame\\
Temperature trends & Five temperature-level values; exact monotonic test when prespecified\\
\bottomrule
\end{tabular}
\end{table}

\section{Persistent-homology definitions and sensitivity}

\subsection{Directed local point clouds}

For center atom $i$ of chemical species $A$, target species $B$, and radius $R$, the periodic point cloud is
\begin{equation}
X_i^{A\rightarrow B}(R)=\{\mathbf 0\}\cup\{\Delta\mathbf r_{ij}^{\rm MIC}:s_j=B,\;\|\Delta\mathbf r_{ij}^{\rm MIC}\|\le R\}.
\end{equation}
The origin is the chosen center. Consequently $A\rightarrow B$ and $B\rightarrow A$ have different centers and generally different diagrams. A Vietoris--Rips complex includes a simplex when all pairwise distances among its vertices do not exceed filtration value $\epsilon$. Finite intervals $[b_k,d_k)$ in dimensions $H_0$, $H_1$, and $H_2$ represent connected-component merging, loops, and cavities. Lifetimes are $\ell_k=d_k-b_k$. Persistence entropy is
\begin{equation}
S_H=-\sum_k p_k\ln p_k,\qquad p_k=\ell_k/\sum_j\ell_j.
\end{equation}
For $H_0$, the lifetime sum is the minimum-spanning-tree length of the point cloud under the adopted metric. For $H_1$ and $H_2$, the same statistic accumulates loop and cavity persistence. Essential intervals with infinite death are counted separately and never assigned an artificial finite lifetime. If a diagram has no positive finite interval, its finite-distribution summaries, total persistence, and entropy are set to zero; this convention distinguishes an empty finite diagram from missing input.

Complete-network diagrams are calculated from a periodic pair-distance matrix. Atomic positions are converted to fractional coordinates, each pair displacement is wrapped independently into $[-1/2,1/2)$, and the wrapped displacement is mapped back through the instantaneous triclinic cell. Ripser is then supplied this minimum-image distance matrix directly. The filtration threshold is the largest pair distance multiplied by $1+10^{-7}$, so every finite interval of the sampled metric is allowed to terminate. Cs and Pb sublattices each contain 64 atoms and are used exhaustively; consequently the headline complete-Cs $H_1$ and complete-Pb $H_1/H_2$ results contain no landmark approximation. I-only and combined Cs--Pb--I networks contain 192 and 320 atoms and are reduced to 64 deterministic farthest-point landmarks. Selection begins with the atom having the smallest summed distance to all other atoms; each subsequent atom is the one farthest from the current selected set. The common cardinality makes their $H_2$ calculations tractable and prevents point count alone from distinguishing chemical scopes.

\begin{table}[h]
\centering
\caption{\label{tab:asph}Primary and sensitivity constructions.}
\begin{tabular}{p{0.22\textwidth}p{0.68\textwidth}}
\toprule
Construction & Definition\\
\midrule
Primary sparse local \ASPH & Three target species for each Cs/Pb/I center, $R=9.5$ \AA, $H_0$--$H_2$\\
Site-resolved \ASPH & Pb-centered immediate-cage channels, $R=8.0$ \AA, five separated 1 ps windows\\
Spatially resolved \ASPH & Fixed-neighbor atom-centered fields retained at every interface site and averaged in 2 \AA\ slabs only for visualization\\
Dopant-network topology & Periodic point clouds containing only substituted Pb sites, $R=8.5$ \AA, with $H_0$ merger scales measuring compactness\\
Cutoff sensitivity & Bulk phase contrasts on matched frames recomputed at $R=8.0$ \AA\\
Global topology & Separate periodic Cs, Pb, and I sublattices and the combined Cs--Pb--I point cloud\\
Coordinate space & Atomic positions with periodic minimum-image distances in \AA\\
Summary family & Count; birth, death, and lifetime extrema and moments; total persistence; persistence entropy\\
Sparse extraction & $10\times25\times320\times3$ target species $=240{,}000$ coordinate-space filtrations\\
\bottomrule
\end{tabular}
\end{table}

The 240,000 atom-centered filtrations follow from 10 state points, 25 production frames per state, 320 atomic centers, and three surrounding species per center. Combining the three possible central species with the three surrounding species yields nine directed chemical channels, although each center contributes only its three species-specific filtrations. All $H_0$--$H_2$ summaries are extracted from each filtration. Across the analyzed NPT frames, the smallest value of half the shortest periodic lattice translation is 9.527 \AA. The primary radius, 9.5 \AA, is the largest value to one decimal place below this minimum and therefore maximizes the sampled radial range while preserving a unique minimum-image neighborhood throughout the 320-atom data. It contains several coordination shells, as required for the medium-range bulk comparison. Recomputing matched frames at 8.0 \AA\ tests whether the reported bulk phase direction depends on this radial extent. The Pb-site calculation uses 8.0 \AA\ for a different physical reason: 159,996 of the 160,000 sampled $\gamma$-phase Pb$\rightarrow$Cs point clouds contain nine points---the central Pb and exactly its eight immediate Cs neighbors---and none contains an additional Cs shell.

For the two local quantities emphasized in the main text, Table~\ref{tab:radius} compares averages over atomic centers on the same 250 state-point--frame combinations at both radii. The rank coefficient measures whether the same frames remain high or low when the radius changes, while phase-sign agreement asks whether the $\gamma-\delta$ difference has the same sign at each of the five temperatures. Individual filtration statistics can depend on radial extent; the table reports the stability of the two quantities used for physical interpretation.

\begin{table}[h]
\centering
\caption{\label{tab:radius}Matched-radius stability of the highlighted local topological quantities.}
\small
\begin{tabular}{lcc}
\toprule
Quantity & Spearman $\rho$, 8.0 vs 9.5 \AA & temperatures with same phase sign\\
\midrule
I$\rightarrow$Pb $H_1$ mean lifetime & 0.861 & five of five\\
Pb$\rightarrow$Cs $H_1$ persistence entropy & 0.840 & five of five\\
\bottomrule
\end{tabular}
\end{table}

\begin{figure}[t]
\centering\includegraphics[width=0.98\textwidth]{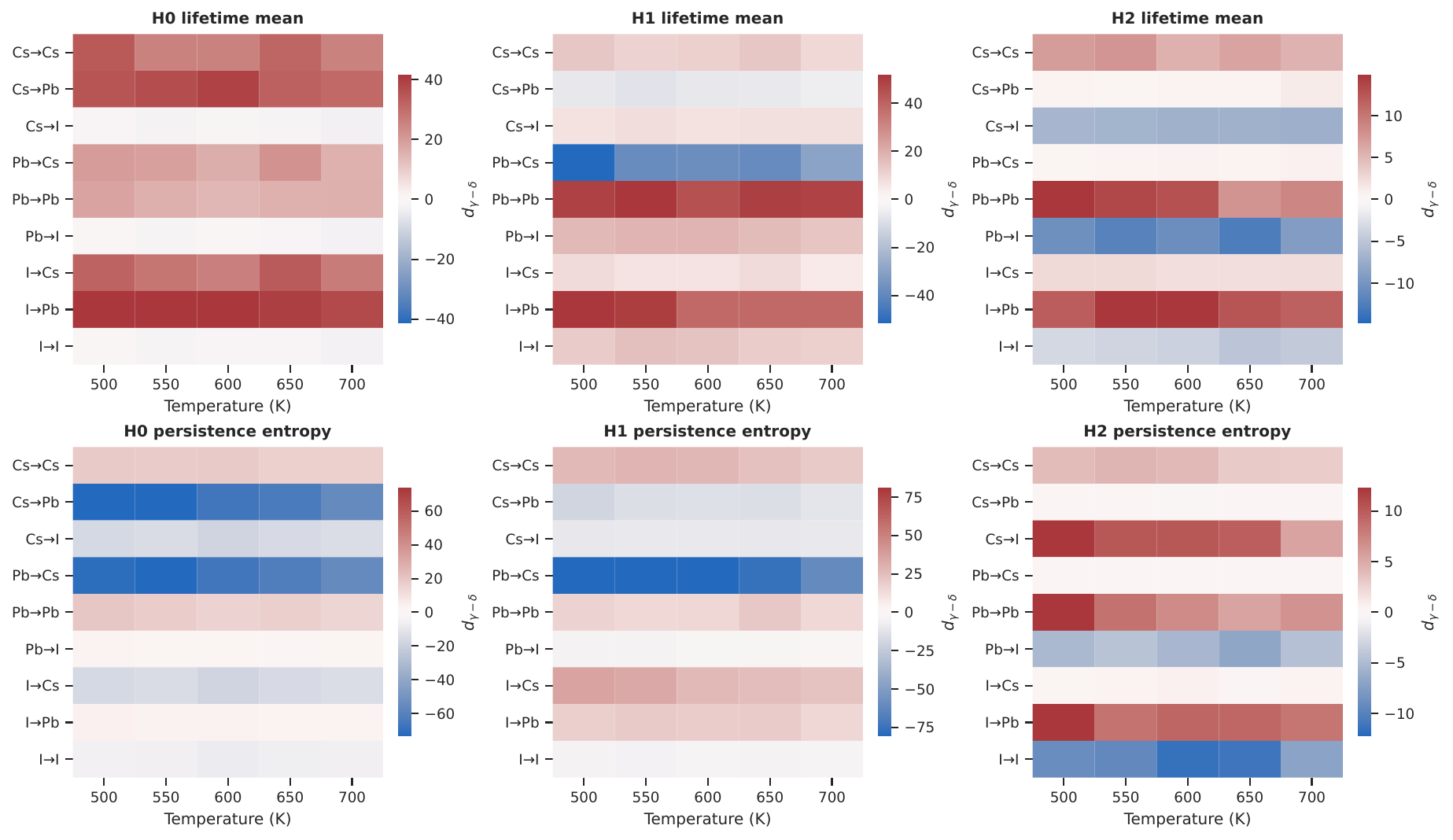}
\caption{\label{fig:s1}Phase contrast for all nine ordered chemical channels, shown for the mean lifetime and persistence entropy in $H_0$, $H_1$, and $H_2$. Values are averaged within each of the five successive portions of the production trajectory; the remaining diagram summaries are retained in the accompanying data tables.}
\end{figure}

\begin{figure}[p]
\centering\includegraphics[width=0.98\textwidth]{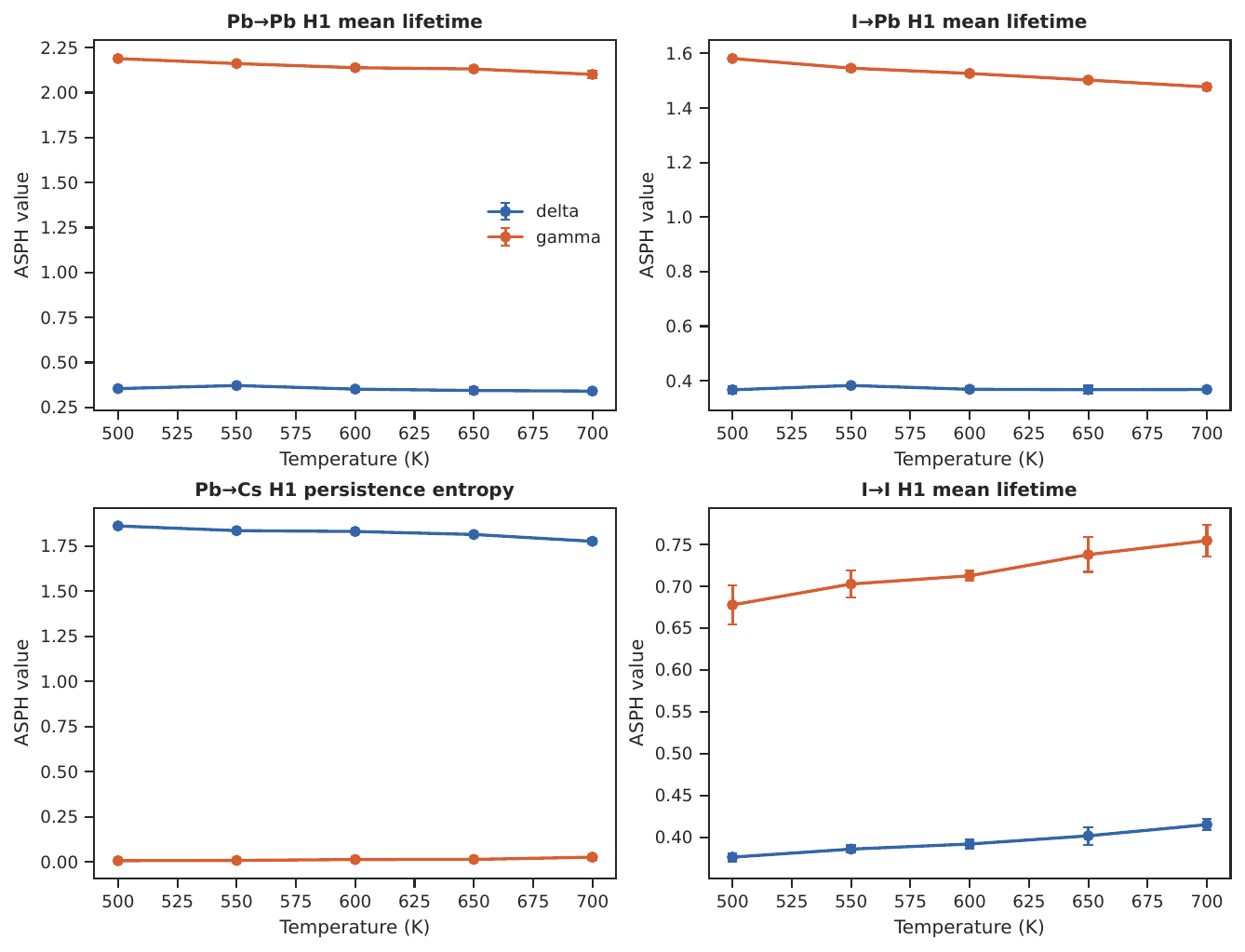}
\caption{\label{fig:s2}Representative local coordinate-space profiles over 500--700 K. Directed channels are plotted separately, emphasizing that exchanging center and target species changes the physical question.}
\end{figure}

\begin{figure}[p]
\centering\includegraphics[width=0.98\textwidth]{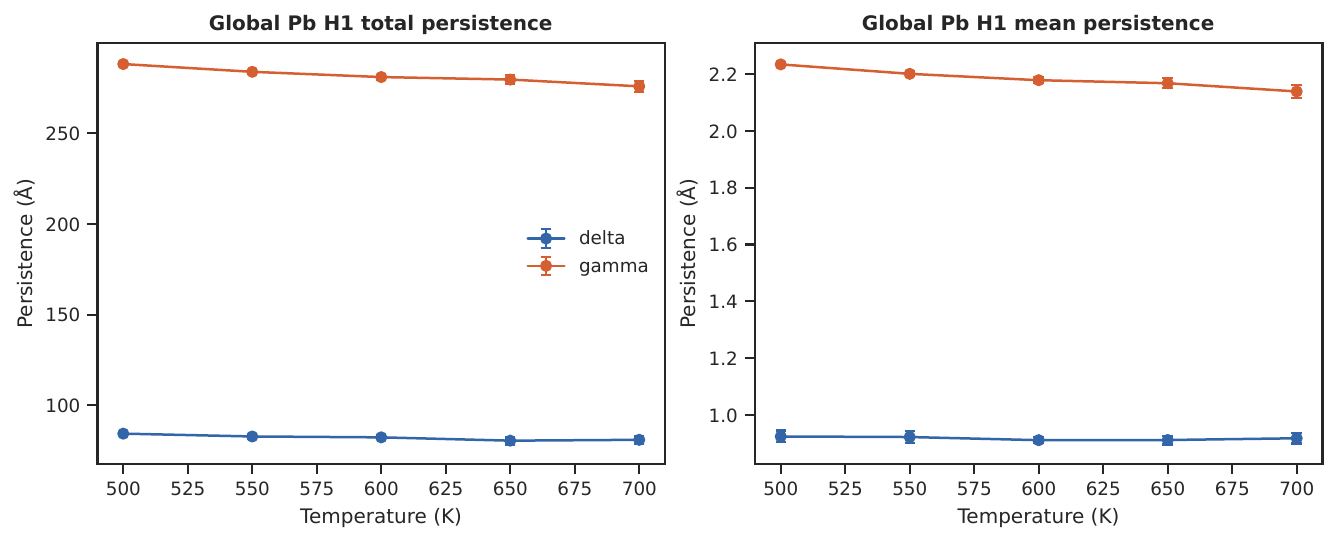}
\caption{\label{fig:s3}Global Pb-network persistence. Total $H_1$ persistence provides the radius-stable medium-range distinction between the connected $\gamma$ network and chain-like $\delta$ network.}
\end{figure}

\begin{figure}[p]
\centering\includegraphics[width=0.98\textwidth]{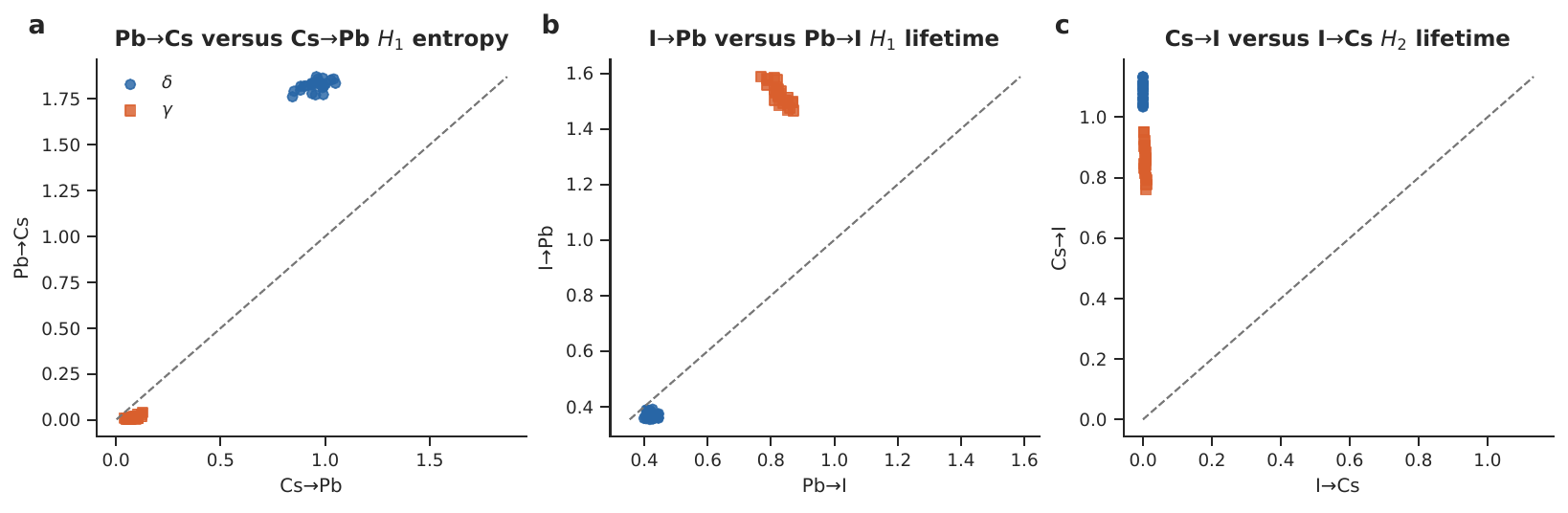}
\caption{\label{fig:s5}Directed-pair nonreciprocity. Each point is one phase--temperature--trajectory-segment estimate. The dashed identity line marks equality between the two directions. Changing $A\rightarrow B$ to $B\rightarrow A$ changes both the central species and the conditioned neighbor species and therefore produces a different physical coordinate.}
\end{figure}

\begin{figure}[p]
\centering\includegraphics[width=0.98\textwidth]{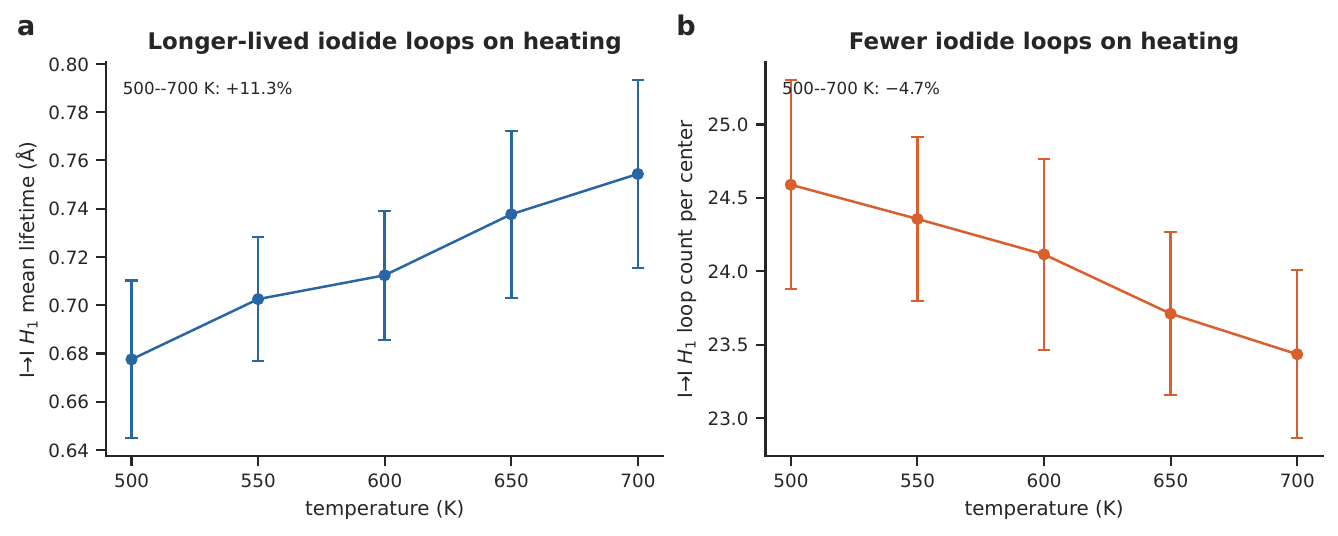}
\caption{\label{fig:s6}Temperature dependence of $\gamma$-phase I$\rightarrow$I loops. (a) Mean $H_1$ lifetime rises by 11.3\% from 500 to 700 K. (b) The mean number of finite loops per iodide center falls by 4.7\%. Error bars are standard deviations over the 25 sampled frames. The permutation probabilities defined in Section S2 are adjusted together for the prespecified temperature trends; the coupled change has $p_{\mathrm{FDR}}=0.049$.}
\end{figure}

\clearpage
\section{Relation to crystallographic geometry}

The reference geometry deliberately covers both the simulation cell and individual PbI$_6$ units. Twenty-seven directly calculated quantities describe cell lengths and angles, volume and anisotropy, coordination and framework connectivity, short and long Pb--I and Cs--I distances, Pb--I--Pb bridges, and Pb off-centering. A further ten properties are calculated for each of the 64 PbI$_6$ units: mean Pb--I bond length, bond-length distortion $\Delta$, bond-angle variance $\sigma^2$, three Pyrovskite distortion or off-centering measures ($\lambda$, $\lambda_2$, and $\lambda_3$), mean cis-angle deviation, mean trans-angle deviation, mean Pb--I--Pb angle, and its deviation from $180^\circ$. Each of these ten per-octahedron properties is represented by seven statistics across the 64 units: mean, standard deviation, minimum, lower quartile, median, upper quartile, and maximum. This contributes $10\times7=70$ values and gives $27+70=97$ geometric quantities in total. Octahedra are assigned by periodic nearest-neighbor geometry; a quantity is omitted for a frame if its assignment is chemically implausible rather than filled with an artificial value.

Tables~\ref{tab:geo27} and \ref{tab:geo70} give the complete construction of the 97-variable basis. Distances are evaluated with periodic minimum images. Pb--I and Cs--I neighbors are defined by 4.0 and 4.5 \AA\ cutoffs, respectively, and an iodide is counted as bridging when it lies within 4.0 \AA\ of exactly two Pb atoms. Pyrovskite quantities retain the definitions of Stanton and Trivedi;\cite{Stanton2023Pyrovskite} the table states explicitly how their per-octahedron values enter the regression.

\begin{longtable}{p{0.30\textwidth}p{0.49\textwidth}p{0.13\textwidth}}
\caption{The 27 cell-, bond-, and coordination-level variables.\label{tab:geo27}}\\
\toprule
Variable(s) & Definition & Unit\\
\midrule
\endfirsthead
\toprule
Variable(s) & Definition & Unit\\
\midrule
\endhead
Volume per atom & instantaneous cell volume divided by 320 & \AA$^3$\\
Mass density & cell mass divided by instantaneous volume & g cm$^{-3}$\\
$a,b,c$ & three instantaneous lattice-vector lengths (three variables) & \AA\\
Cell-length spread & population standard deviation of $a,b,c$ & \AA\\
Maximum angle deviation & largest absolute deviation of a cell angle from $90^\circ$ & degree\\
Pb--I distance mean, spread, minimum, maximum & four statistics over assigned periodic Pb--I neighbors & \AA\\
Cs--I distance mean, spread, minimum, maximum & four statistics over assigned periodic Cs--I neighbors & \AA\\
Mean Pb and Cs coordination & mean assigned I-neighbor count for each cation species (two variables) & count\\
Mean Pb and Cs off-centering & mean cation displacement from the centroid of its six or twelve nearest iodides, respectively (two variables) & \AA\\
Mean PbI$_6$ bond distortion & mean dimensionless bond-length distortion over assigned octahedra & dimensionless\\
Mean cis and trans I--Pb--I angles & means over assigned cis and trans angle sets (two variables) & degree\\
Bridging-I fraction & fraction of iodides assigned to two Pb centers & fraction\\
Pb--I--Pb angle mean, spread, minimum, maximum & four statistics over assigned bridging iodides & degree\\
\bottomrule
\end{longtable}

\begin{longtable}{p{0.28\textwidth}p{0.47\textwidth}p{0.17\textwidth}}
\caption{The ten per-octahedron properties that generate the remaining 70 variables. Each row contributes seven values: mean, population standard deviation, minimum, 25th percentile, median, 75th percentile, and maximum over the assigned PbI$_6$ units.\label{tab:geo70}}\\
\toprule
Per-octahedron property & Definition & Unit\\
\midrule
\endfirsthead
\toprule
Per-octahedron property & Definition & Unit\\
\midrule
\endhead
Pb--I distance & six-bond mean returned by Pyrovskite & \AA\\
Bond distortion $\Delta$ & normalized variance of the six Pb--I distances & dimensionless\\
Bond-angle variance $\sigma^2$ & variance of the twelve cis I--Pb--I angles about $90^\circ$ & degree$^2$\\
$\lambda$ & product-to-maximum ratio of the three scaled pairwise off-centering components defined by Pyrovskite & \AA$^2$\\
$\lambda_2$ & maximum of the three scaled pairwise off-centering components & \AA\\
$\lambda_3$ & product of the three scaled pairwise off-centering components & \AA$^3$\\
Cis I--Pb--I angle & individual assigned cis angles & degree\\
Trans I--Pb--I angle & individual assigned trans angles & degree\\
Pb--I--Pb angle & bridge angle returned for the connected octahedral network & degree\\
Pb--I--Pb deviation & absolute deviation of the bridge angle from $180^\circ$ & degree\\
\bottomrule
\end{longtable}

Three topological inputs are compared: the collection of atom-centered ordered-pair summaries, the complete-sublattice summaries, and their union. To test transfer in temperature, both phases at one temperature are removed, a ridge-regression model is fitted to the other four temperatures, and the removed temperature is predicted. Ridge regression is a linear model with a penalty on large squared coefficients, which stabilizes fitting when predictors are correlated. Repeating the temperature exclusion five times gives one prediction for every temperature without ever training on frames from the temperature being predicted. A separate, more demanding test asks whether topology reproduces small fluctuations within one phase--temperature trajectory: nonoverlapping temporal portions from that same trajectory are used for training and testing. The first test measures recovery of phase and thermal-state geometry; the second measures instantaneous vibrational detail. The ridge penalty is chosen using only the training temperatures or temporal portions.

The reconstruction has direct structural content. I$\rightarrow$Cs $H_0$ connectivity follows the Cs--I distance ($\rho=0.77$); I$\rightarrow$Pb $H_0$ follows the short Pb--I distance ($\rho=0.72$); the scale at which I$\rightarrow$I loops appear follows the Pb--I distance ($\rho=0.65$); Cs$\rightarrow$I cavity persistence decreases with angular disorder ($\rho=-0.63$ and $-0.61$ for two measures); and Pb$\rightarrow$I loop persistence follows the Pb--I--Pb angle ($\rho=0.59$). These relationships assign chemical meaning to representative filtration scales. The remaining topological variation describes how these local units are organized across the larger chemical network.

To identify information not reproduced by these 97 quantities, each topological feature is predicted from the complete geometry basis using the same temperature-transfer procedure. For the excluded temperature, the geometry-based prediction is subtracted from the observed topological value. This difference is the geometry-conditioned topological signal: it is the part of topology left after the measured cell and octahedral geometry has been accounted for by the stated linear model. The standard deviation of complete-sublattice Cs $H_1$ closure scales and the maximum lifetime of complete-sublattice Pb $H_2$ retain the same phase direction in every successive trajectory segment at all five temperatures, giving 25 comparisons. To compare quantities with different units, the $\gamma-\delta$ difference is divided by the pooled standard deviation of the two phases' five trajectory-segment estimates; this dimensionless difference is denoted $d$. The smallest magnitudes of $d$ are 1.70 and 1.92, respectively, both at 550 K. The Cs quantity measures heterogeneity in the distances at which network loops close. The Pb quantity measures the longest-lived enclosed arrangement in the filtration and should not be interpreted as the radius of a literal pore. Both retain how local units are organized over the larger species network, information not present in distributions of individual-octahedron distortions.

\begin{table}[h]
\centering
\caption{\label{tab:geometry}Geometry--topology validation summary. Here $d$ is the difference between phase means divided by their pooled standard deviation.}
\small
\begin{tabular}{lcc}
\toprule
Test & quantity & result\\
\midrule
Temperature-transfer reconstruction & median $R^2$, local \ASPH & 0.68\\
Temperature-transfer reconstruction & median $R^2$, all \ASPH & 0.66\\
Within-trajectory reconstruction & median $R^2$ & 0.07\\
Geometry-conditioned global Cs $H_1$ closure spread & smallest temperature-level $|d|$ & 1.70 (550 K)\\
Geometry-conditioned global Pb $H_2$ maximum lifetime & smallest temperature-level $|d|$ & 1.92 (550 K)\\
\bottomrule
\end{tabular}
\end{table}

\begin{figure}[p]
\centering\includegraphics[width=0.98\textwidth]{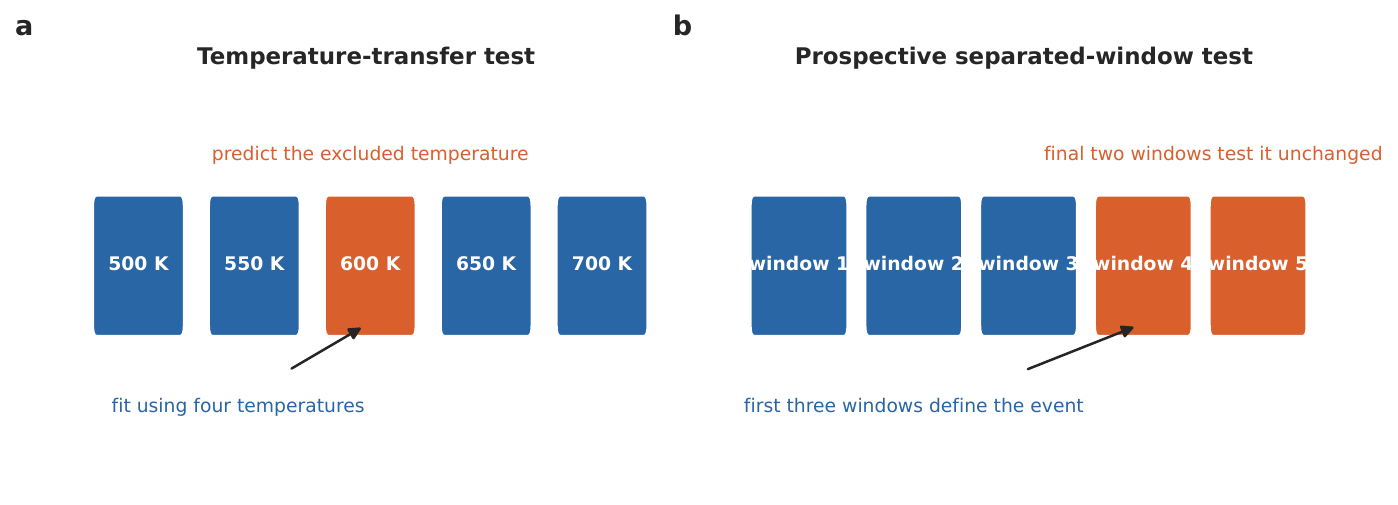}
\caption{\label{fig:s7}Validation design. (a) Temperature transfer excludes both phases at one temperature, fits on the other four temperatures, and predicts the excluded data. The orange position is rotated through all five temperatures. (b) The prospective atom-centered analysis uses the first three separated 1 ps windows to define an event and the final two windows to test it without changing the definition.}
\end{figure}

\begin{figure}[p]
\centering\includegraphics[width=0.98\textwidth]{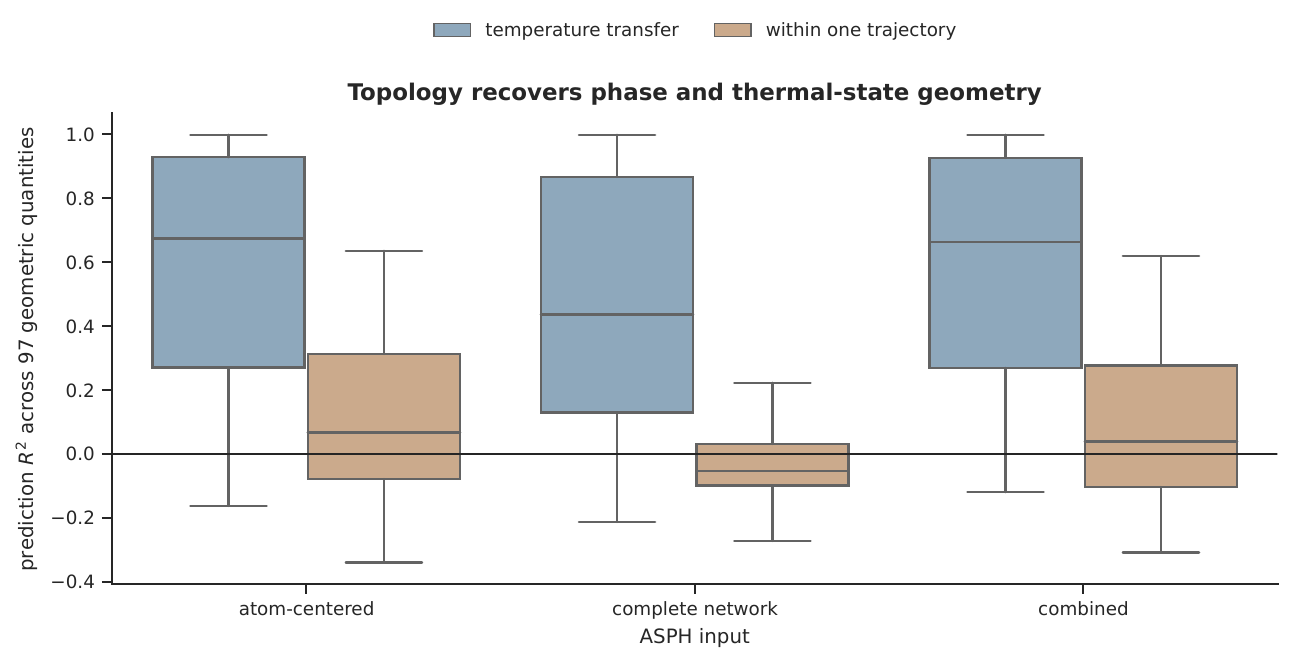}
\caption{\label{fig:s8}Prediction of individual geometric quantities from atom-centered, complete-sublattice, and combined \ASPH. In each cycle, both phases at the plotted temperature are excluded while a separate ridge regression for each geometric quantity is fitted to the other four temperatures.}
\end{figure}

\begin{figure}[p]
\centering\includegraphics[width=0.98\textwidth]{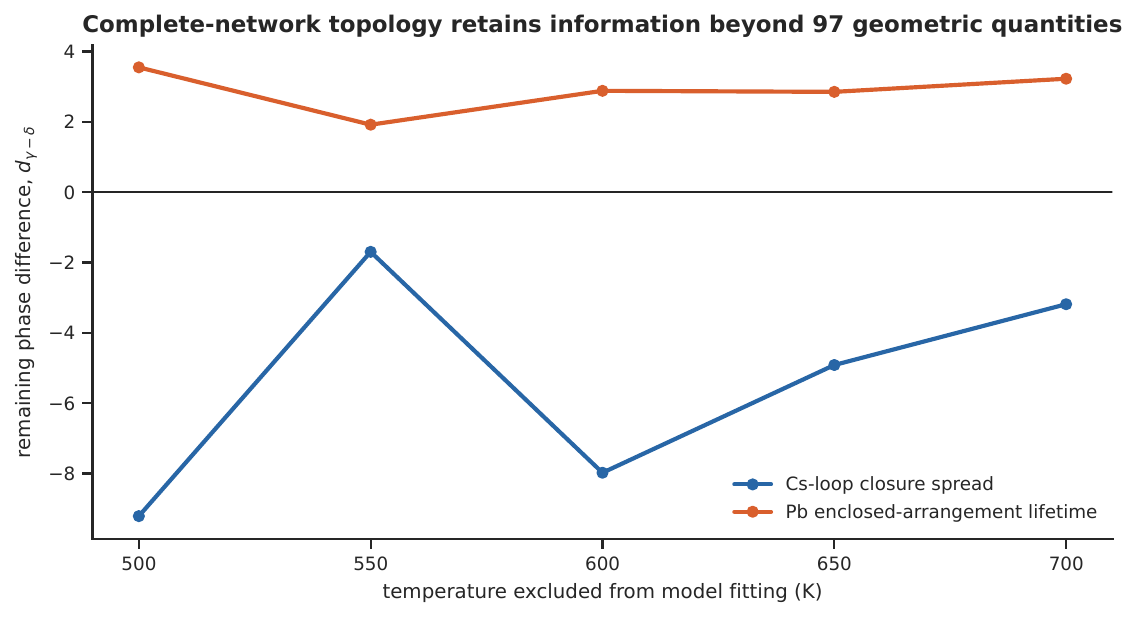}
\caption{\label{fig:s9}Phase differences in topology after removing the part predicted by all 97 geometric quantities. The two selected complete-sublattice features retain the same phase direction at every excluded temperature.}
\end{figure}

\begin{figure}[p]
\centering\includegraphics[width=0.98\textwidth]{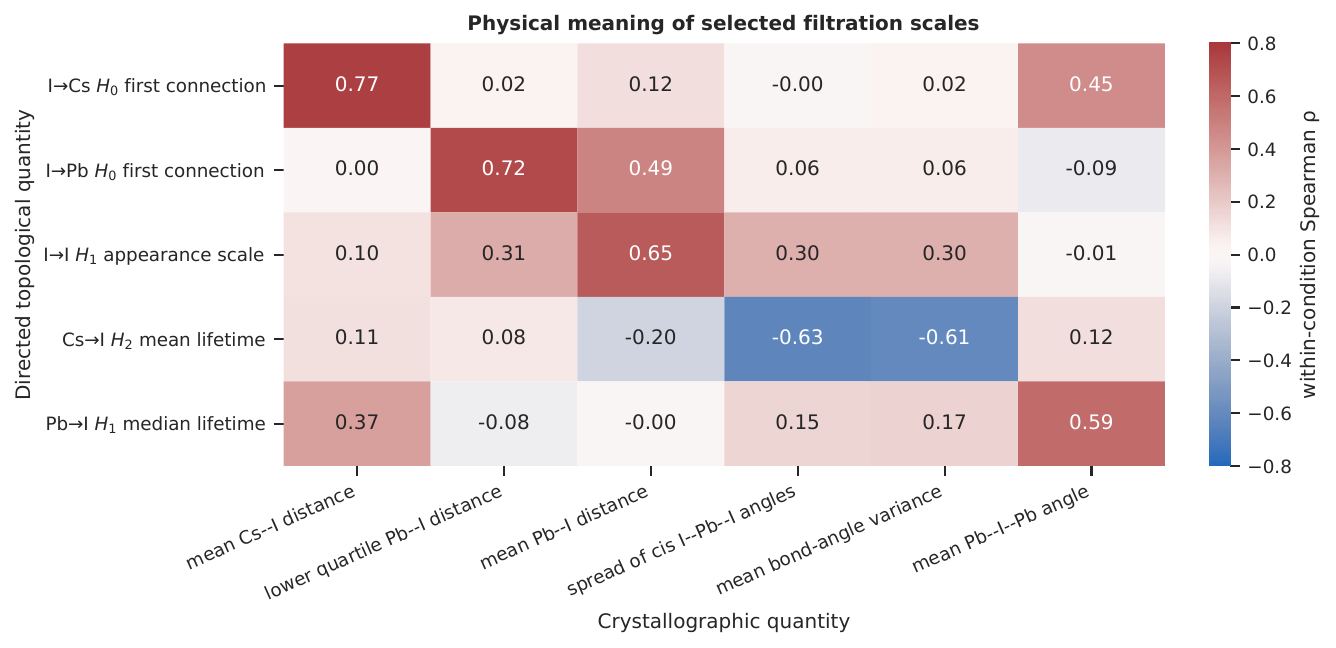}
\caption{\label{fig:s10}Physical map between representative \ASPH\ channels and octahedral/cell geometry. Correlations support interpretation but do not establish that the descriptor families are identical.}
\end{figure}

\begin{figure}[p]
\centering\includegraphics[width=0.98\textwidth]{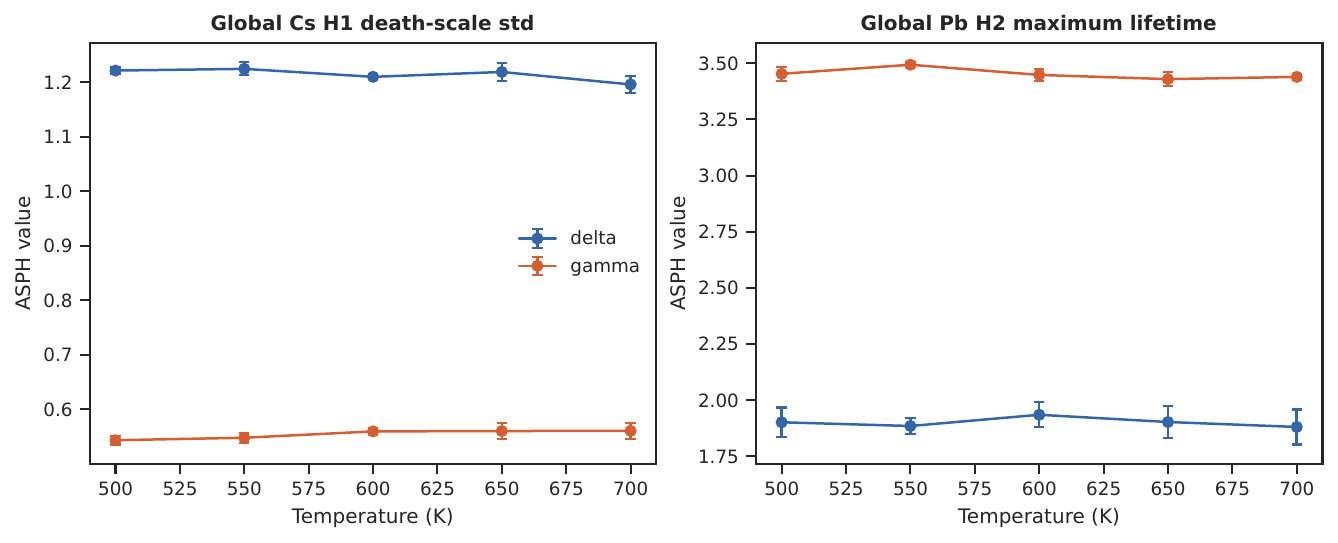}
\caption{\label{fig:s11}Profiles of nonredundant global topology features retained after correlation-based feature reduction.}
\end{figure}

\clearpage
\section{Prospective test of rare atom-centered environments}

The atom-centered analysis uses Pb because Pb is the framework node whose cage and connectivity must reorganize during the phase transformation. At each temperature, 8.0 \AA\ neighborhoods are evaluated in five separated 1 ps windows: production frames 0--99, 1575--1674, 3150--3249, 4725--4824, and 6300--6399. Of the 160,000 $\gamma$-phase Pb$\rightarrow$Cs point clouds, 159,996 contain nine points---the central Pb and exactly the eight Cs sites of its immediate A-site cage---and none reaches an additional Cs shell. The calculation evaluates nine directed center--target pairs. For each pair, 18 nonconstant quantities describe $H_0$--$H_2$ feature counts, lifetimes, total persistence, and persistence entropy; this gives $9\times18=162$ possible local coordinates. Only the first three windows are used to retain coordinates whose phase difference has the same direction at all five temperatures and is at least 1.5 pooled standard deviations at every temperature. Forty-nine coordinates satisfy both requirements, and their numerical upper-tail thresholds are then fixed. The final two windows are evaluated without changing the coordinates or thresholds. Across five temperatures this gives $5\times2=10$ later-window comparisons. They measure recurrence at separated times within the available trajectories, not ten independent simulations. Because the response extends 0.5 ps beyond the event time, frames 6300--6349 support the future-displacement response in the final window; frames 6350--6399 remain available only for quantities evaluated at the event time and for the spatial-association analysis.

A $\delta$-range event is an unusually large Pb$\rightarrow$Cs $H_1$ value in $\gamma$ that falls inside the central 80\% of the $\delta$ distribution determined from the first three windows. It therefore denotes overlap in this particular Cs-cage connectivity scale. When an event occurs at time $t$, the response is the Pb displacement between $t$ and $t+0.5$ ps after subtracting the mass-weighted mean displacement of all 320 atoms. This mass-weighted subtraction removes rigid translation of the complete simulation cell while retaining motion of Pb relative to the other atoms. For each Pb atom, displacement after its event environments is compared with displacement after its ordinary environments; these within-atom differences are then summarized for each of the ten later windows. This is the primary, unadjusted effect. The two-sided Wilcoxon signed-rank test orders the absolute values of the ten paired window effects, retains their signs, and tests whether positive and negative effects are balanced around zero. The 49 retained topological channels are paired with the force, speed, coordination, distortion, off-centering, and future-displacement responses available for their central species. Excluding combinations with usable data in fewer than eight later-window comparisons leaves 313 channel--response tests. Their probabilities form one prespecified evaluation family and are adjusted together by the Benjamini--Hochberg procedure.

A second analysis includes quantities available at time $t$: instantaneous Pb speed, force magnitude, Pb off-centering, PbI$_6$ bond distortion, the number of Cs atoms in the point cloud, its radial extent, Pb identity, and frame identity. The reported coefficient is the remaining event-associated displacement after these simultaneous variables are fitted. Its windowwise median measures the stability of the effect to this adjustment; $p_{\mathrm{FDR}}$ belongs to the primary within-atom contrast defined above.

\begin{table}[h]
\centering
\caption{\label{tab:precursor}Later-time results for the Pb$\rightarrow$Cs $H_1$ events defined using only the first three separated windows. The Benjamini--Hochberg-adjusted probabilities apply only to the unadjusted within-atom contrasts.}
\small
\begin{tabular}{lccc}
\toprule
Feature & raw effect (\AA) & adjusted median (\AA) & sign consistency\\
\midrule
$H_1$ lifetime sum & 0.172 ($p_{\rm FDR}=0.038$) & 0.162 & unadjusted: all ten; adjusted: eight of ten\\
$H_1$ maximum lifetime & 0.125 ($p_{\rm FDR}=0.038$) & 0.142 & unadjusted: all ten; adjusted: seven of ten\\
\bottomrule
\end{tabular}
\end{table}

The rare-event support is given explicitly in Table~\ref{tab:precursor-support}. ``Observations'' counts Pb--frame events with a complete 0.5 ps future displacement; ``sites'' counts distinct Pb atoms contributing at least one such event. Repeated events at one site are averaged before the within-atom contrast is formed.

\begin{table}[h]
\centering
\caption{\label{tab:precursor-support}Pb--frame observations and distinct Pb sites supporting each future-displacement comparison. W3 and W4 denote the fourth and fifth separated windows, which are the two windows reserved for evaluation.}
\small
\begin{tabular}{ccrrrr}
\toprule
$T$ (K) & window & \multicolumn{2}{c}{$H_1$ lifetime sum} & \multicolumn{2}{c}{$H_1$ maximum lifetime}\\
 & & observations & sites & observations & sites\\
\midrule
500 & W3 & 77 & 5 & 109 & 8\\
500 & W4 & 7 & 1 & 12 & 2\\
550 & W3 & 56 & 4 & 62 & 4\\
550 & W4 & 15 & 2 & 21 & 3\\
600 & W3 & 35 & 5 & 37 & 9\\
600 & W4 & 41 & 7 & 26 & 4\\
650 & W3 & 79 & 8 & 100 & 7\\
650 & W4 & 7 & 1 & 4 & 2\\
700 & W3 & 70 & 6 & 52 & 3\\
700 & W4 & 68 & 4 & 33 & 4\\
\bottomrule
\end{tabular}
\end{table}

Spatial coherence is evaluated for the continuous Pb$\rightarrow$Cs persistence value $x_i$ on a periodic graph joining every Pb site to its six nearest Pb neighbors. With $w_{ij}=1$ for an edge, $N=64$ sites, and $W=\sum_{ij}w_{ij}$ directed neighbor links, Moran's statistic is
\begin{equation}
I=\frac{N}{W}\frac{\sum_{ij}w_{ij}(x_i-\bar{x})(x_j-\bar{x})}{\sum_i(x_i-\bar{x})^2}.
\end{equation}
For randomly permuted values on the same graph, its expectation is $-1/(N-1)$. We therefore plot $I+1/(N-1)$: zero denotes the random-placement expectation, a positive value means neighboring Pb sites tend to have similar persistence, and a negative value means neighboring values tend to differ. The median excess is 0.012 for both reported $H_1$ quantities. Total persistence is positive in all ten later windows and maximum lifetime in nine. Adjacent radius-8 \AA\ environments share atoms, so the small positive value describes weak local smoothness of the continuous coordinate.

\begin{figure}[p]
\centering\includegraphics[width=0.98\textwidth]{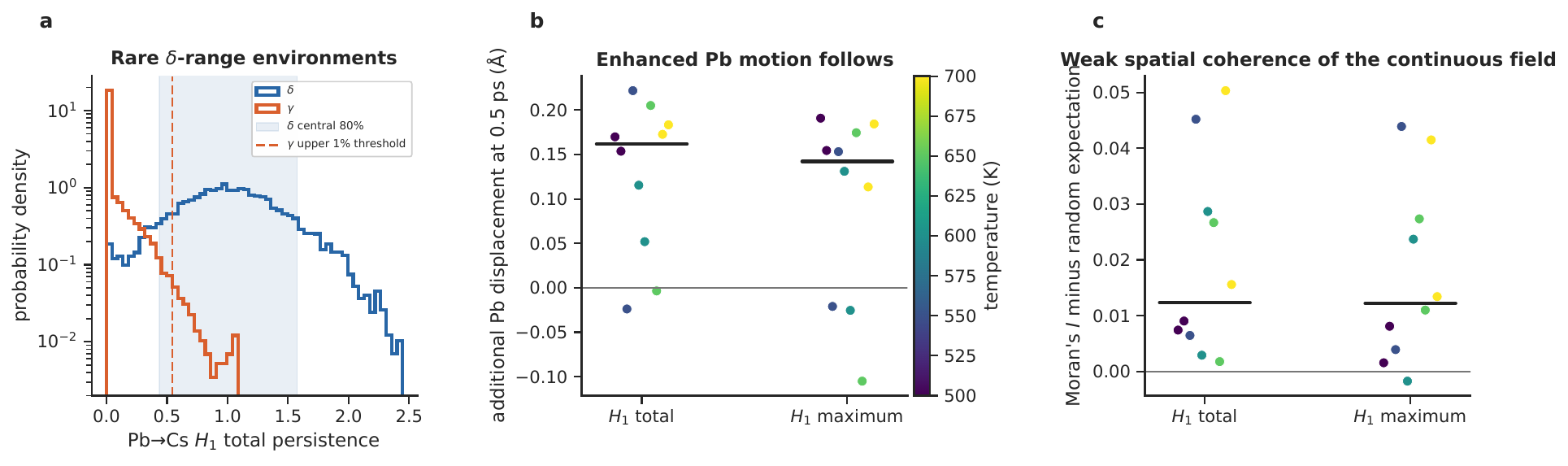}
\caption{\label{fig:s12}Prospective site-resolved analysis. The first three separated 1 ps windows define the $\delta$-range tail; the final two windows at each temperature test subsequent Pb displacement and the nearest-neighbor coherence of the continuous persistence field without redefining the event. Panel c reports Moran's $I$ relative to its random-placement expectation, as defined above.}
\end{figure}

\clearpage
\section{Interface construction and descriptor comparison}

\subsection{Interface construction}

The two parent structures are complete 320-atom cells taken from the 600 K production trajectories: $\delta$ frame 3866 at 48.66 ps and $\gamma$ frame 1391 at 23.91 ps. They are representative thermal structures rather than ideal crystals. Five quantities are calculated for every production frame: volume per atom, DFT potential energy per atom, root-mean-square force, pressure, and temperature. For each quantity, the difference from the trajectory median is divided by $1.4826$ times its median absolute deviation; the factor 1.4826 places this robust spread estimate on the same scale as a standard deviation for normally distributed data. The selected frame minimizes the sum of the five squared dimensionless deviations and is consequently close to the trajectory median in all five physical quantities rather than exceptional in any one of them. Frame selection therefore depends only on these thermodynamic and force quantities. The complete cells are retained without reconstructing a primitive cell. The interface normal uses parent axes 1 and 0 for $\delta$ and $\gamma$, respectively. A search over small integer in-plane repeats identifies the following commensurate pair:
\begin{equation}
M_{\delta}=\begin{pmatrix}0&-1\\2&0\end{pmatrix},\qquad
M_{\gamma}=\begin{pmatrix}-1&-1\\1&-1\end{pmatrix}.
\end{equation}
The common in-plane lattice is chosen midway between the two repeated cells so that the mismatch is shared rather than imposed entirely on one phase. The resulting Green--Lagrange strain, a finite-deformation measure that is zero for an unstrained lattice, is
\begin{equation}
E_{\delta}=\begin{pmatrix}-0.01128&-0.00364\\-0.00364&-0.02921\end{pmatrix},\quad
E_{\gamma}=\begin{pmatrix}0.01173&0.00421\\0.00421&0.03202\end{pmatrix},
\end{equation}
with principal strains $(-2.99,-1.06)\%$ and $(1.09,3.29)\%$. These values keep the largest imposed strain below 3.3\%. The 2560-atom cell vectors (in \AA) are
\begin{equation}
\begin{aligned}
\mathbf a&=(-0.0456,\ 0.3391,\ -36.1242),\\
\mathbf b&=(37.8246,\ 0.2967,\ 0.4852),\\
\mathbf c&=(0.7958,\ -99.9264,\ -0.9391).
\end{aligned}
\end{equation}
Periodic stacking necessarily produces two boundaries. Their initial fractional coordinates along the interface normal are 0.0056 and 0.4611. The area of each boundary is 1366 \AA$^2$ and the cell is 99.9 \AA\ along its normal; each phase contributes 1280 atoms. The distances between successive boundaries give $\delta$ and $\gamma$ slab thicknesses of 43.276 and 50.485 \AA, respectively.

The relative lateral translation of the two phases is not fixed by lattice matching. Nine alignments on a $3\times3$ fractional grid were therefore compared by single-point energy. The three lowest-energy alignments, $(2/3,0)$, $(2/3,1/3)$, and $(2/3,2/3)$, received the same 120-step fixed-cell relaxation. Registry $(2/3,0)$ had the lowest relaxed energy and was selected before descriptor profiles were calculated.

The machine-learned interatomic potential starts from MACE-MH-0 with the energy head pretrained on MatPES r2SCAN calculations.\cite{Batatia2022MACE,Batatia2025CrossLearning} The weights are fitted to 1110 bulk AIMD structures spanning both phases and all five temperatures. A further 400 structures from separate contiguous time intervals are used to compare learning rates and the relative weights assigned to energy, force, and stress errors. The selected settings are 50 epochs, batch size 2, energy/force/stress loss weights 100/100/1000, learning rate $10^{-5}$, and random seed 1729. After these choices are fixed, performance is evaluated once on a third set of 400 structures taken from different trajectory intervals. No weights or settings are updated from this final set. The relative-energy mean absolute error on it, after applying one global energy offset fitted to the 1110 training structures, is 8.9 meV atom$^{-1}$; the median force root-mean-square error is 0.014 eV \AA$^{-1}$ (0.018 eV \AA$^{-1}$ in the least accurate condition); the error in the $\delta$--$\gamma$ energy difference is 0.6 meV atom$^{-1}$; and the median stress root-mean-square error is 0.12 GPa. After interface relaxation, the root-mean-square and maximum MACE forces are 0.003 and 0.033 eV \AA$^{-1}$. Langevin dynamics at fixed particle number, volume, and temperature (NVT) uses 600 K, a 1 fs step, 5 ps equilibration, and 20 ps production sampled every 20 fs, giving $(20\ \mathrm{ps})/(0.020\ \mathrm{ps})+1=1001$ saved production frames including both endpoints. The production trajectory begins from registry $(2/3,0)$ defined above. Descriptor profiles are calculated along this trajectory.

\subsection{Calculation of comparison fields}

All comparators are calculated on the same interface frames and reduced to a spatially resolved scalar field before slab averaging. Standard SOAP neighbor densities contain Cs, Pb, and I, with $r_{\rm cut}=6$ \AA\ and radial and angular expansion orders $n_{\max}=l_{\max}=6$; these settings produce a 1197-component vector at each center. In every frame separately, principal-component analysis is fitted without phase labels using a fixed randomized-decomposition seed. Its first component---the linear combination with the greatest variance among all 2560 site vectors in that frame---provides the scalar SOAP field. Steinhardt $q_6$ measures the degree of local sixth-order bond-orientational order using neighbors within 4.5 \AA. Ten Pyrovskite measures are evaluated for each Pb site: Pb--I bond mean, standard deviation, and range; bond-length distortion $\Delta$; bond-angle variance $\sigma^2$; the three $\lambda$ off-centering measures; and mean cis- and trans-angle deviations. The cis-angle deviation has the highest median sigmoid-profile $R^2$ (0.281, compared with 0.232 for the next-best trans-angle deviation) and represents the best Pyrovskite result in the main comparison. The topological coordinate is the mean lifetime of Pb$\rightarrow$Pb $H_1$ loops formed from the 16 nearest Pb atoms; 8 and 12 neighbors provide sensitivity tests.

For each frame, the slab profile is fitted to $f(z)=a+b/[1+\exp(-(z-z_0)/w)]$, where $z_0$ is the boundary position and $w$ controls its width. The coefficient of determination $R^2=1-\sum_i[x_i-f(z_i)]^2/\sum_i(x_i-\bar{x})^2$ measures the fraction of spatial variation reproduced by this single transition. Bulk phase ordering is measured by the area under the receiver-operating-characteristic curve (AUC), equivalently the probability that a randomly chosen site from one phase is ranked above a randomly chosen site from the other after orienting the descriptor. Interface--interior differences use Cliff's $\delta=P(x_{\rm interface}>x_{\rm interior})-P(x_{\rm interface}<x_{\rm interior})$; $-1$ therefore means that every sampled interface value is smaller than every corresponding interior value.

Each atom retains the $\delta$- or $\gamma$-parent identity of the slab from which it was taken. Boundary positions are defined from these parent identities, independently of every descriptor being compared. At each frame, every point on a 720-bin periodic grid along the stacking direction is assigned the majority parent identity among its 32 nearest atoms; a five-bin circular average removes isolated label changes, and the two 50\% crossings define the boundary planes. Cartesian distance normal to these planes is calculated from the reciprocal-lattice normal, so cell shear and thermal strain are retained. The signed distance is negative on the constructed $\delta$ side and positive on the $\gamma$ side. For visualization, scalar values are averaged in 2 \AA\ slabs. Within each frame, a plotted descriptor $x$ is converted to $s(x-\widetilde{x})/\mathrm{IQR}(x)$, where $\widetilde{x}$ is the median, $\mathrm{IQR}$ is the interquartile range spanning the central 50\% of values, and $s=\pm1$ orients the two interior bulk medians from $\delta$ to $\gamma$. This normalization changes neither spatial ordering nor the profile fit.

The resolution benchmark fits the raw atomwise values in each selected frame to
\begin{equation}
f(z)=f_-+\frac{f_+-f_-}{1+\exp[-(z-z_0)/w]}.
\end{equation}
The reported profile fidelity is $R^2=1-\sum_j[f_j-f(z_j)]^2/\sum_j(f_j-\bar f)^2$; the fitted center is $z_0$ and the 10--90\% width is $4.394|w|$. Twenty-five evenly spaced frames provide the matched ASPH, SOAP, Steinhardt, and Pyrovskite comparison. After the Pb$\rightarrow$Pb $H_1$ channel and 16-neighbor construction are fixed, 101 frames provide the higher-time-resolution evaluation and PDynA-coverage analysis. The parent-phase identities define the reference boundary planes used to measure profile fidelity, width, and interface--interior differences.

The Pb-layer density normal to the interface fixes the regions used for the motif-coverage and interface--interior comparisons. Over the 101 analyzed frames, a 4 \AA\ interval contains exactly 32 $\gamma$-origin Pb atoms---16 in the boundary-facing layer at each periodic interface---and the next $\gamma$ Pb layer starts no closer than 4.6 \AA. The identical normal-distance interval is applied on the $\delta$ side. An interior margin of 9.6 \AA\ excludes the three closest $\gamma$ Pb layers at each boundary and leaves a bulk-like plateau; the directions of the descriptor comparisons are unchanged at 8.0 and 11.2 \AA\ margins. Within the 4 \AA\ interval, 40\% of Pb sites on the $\delta$ side and 7\% on the $\gamma$ side admit the ideal-octahedron PDynA fit; the corresponding 9.6 \AA\ interior fractions are 100\% and 84\%. The octahedral profile therefore loses coverage where distortion is largest, whereas Pb$\rightarrow$Pb topology remains defined and measures the reorganized network without assigning an ideal PbI$_6$ unit.

\begin{table}[h]
\centering
\caption{\label{tab:interface}Interface benchmark.}
\begin{tabular}{lcc}
\toprule
Descriptor & median profile $R^2$ & minimum profile $R^2$\\
\midrule
Pb$\rightarrow$Pb $H_1$ \ASPH & 0.963 & 0.953\\
Pyrovskite cis-angle deviation & 0.281 & 0.226\\
Steinhardt $q_6$ & 0.221 & 0.193\\
Unsupervised SOAP PC1 & 0.004 & 0.003\\
\bottomrule
\end{tabular}
\end{table}

All nine directed species pairs were evaluated before the spatial interpretation was restricted to the Pb framework. Table~\ref{tab:interface-pairs} compares the same fixed-16 $H_1$ lifetime mean for every pair, thereby avoiding selection among different homology dimensions or summary statistics. The phase AUC is the probability of correctly ordering one site from each bulk-like region. The boundary contrast is calculated separately on each side as the absolute difference between the mean value within 4 \AA\ of the boundary and the interior mean, divided by the interior standard deviation; the two sides are averaged within a frame and the table reports the temporal median. Pb$\rightarrow$Pb combines exact phase ordering with the largest boundary contrast and directly represents the framework whose chain and corner-sharing connectivity differs between $\delta$ and $\gamma$.

\begin{table}[h]
\centering
\caption{All directed fixed-16 $H_1$ lifetime-mean fields in the 25-frame interface comparison.\label{tab:interface-pairs}}
\small
\begin{tabular}{lccc}
\toprule
Directed channel & median phase AUC & minimum phase AUC & median boundary contrast\\
\midrule
Cs$\rightarrow$Cs & 0.999 & 0.998 & 2.00\\
Cs$\rightarrow$I  & 0.815 & 0.754 & 1.98\\
Cs$\rightarrow$Pb & 1.000 & 0.992 & 2.95\\
I$\rightarrow$Cs  & 0.941 & 0.913 & 1.01\\
I$\rightarrow$I   & 0.975 & 0.952 & 1.19\\
I$\rightarrow$Pb  & 1.000 & 1.000 & 2.62\\
Pb$\rightarrow$Cs & 0.909 & 0.871 & 1.05\\
Pb$\rightarrow$I  & 0.922 & 0.868 & 1.39\\
Pb$\rightarrow$Pb & 1.000 & 1.000 & 5.23\\
\bottomrule
\end{tabular}
\end{table}

\begin{figure}[t]
\centering\includegraphics[width=0.98\textwidth]{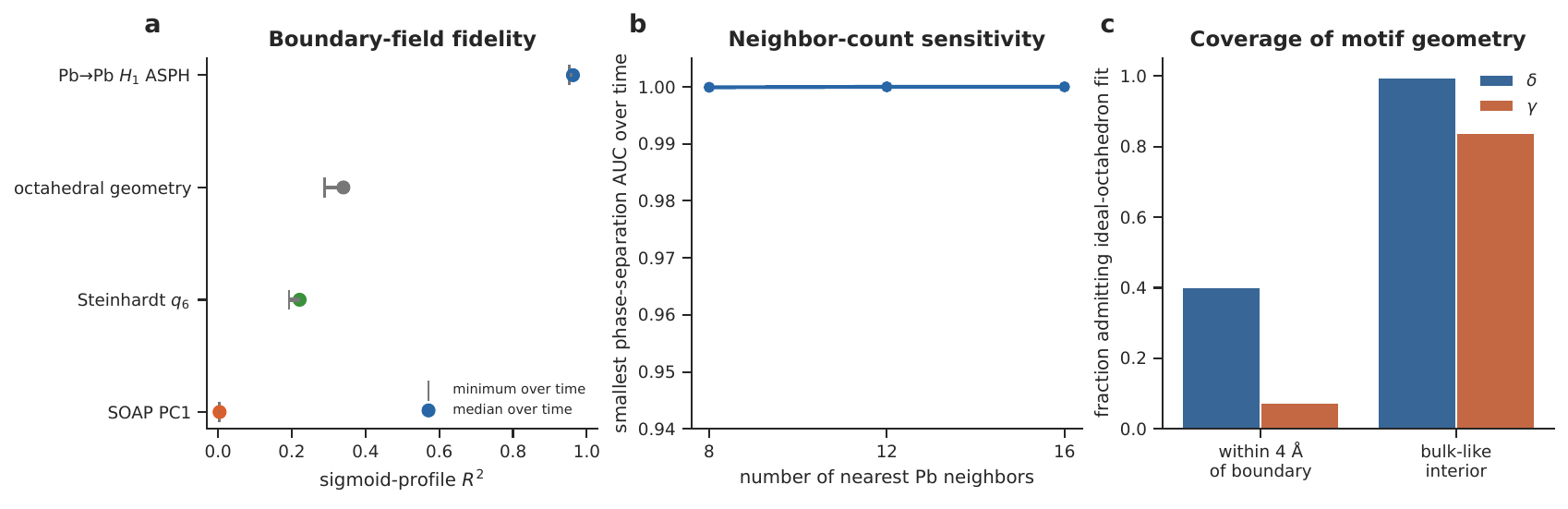}
\caption{\label{fig:s13}Interface-descriptor checks. (a) Median and minimum fidelity of a single sigmoidal boundary profile throughout the trajectory for the direct scalar fields. SOAP is represented by the first unsupervised principal component of the vectors defined above. (b) Phase separation from Pb$\rightarrow$Pb $H_1$ remains effectively exact for neighborhoods containing 8, 12, or 16 nearest Pb atoms. AUC is the probability that a randomly chosen site from one bulk phase is ordered correctly relative to one from the other phase. (c) Fraction of Pb sites for which PDynA can assign the ideal-octahedron fit.}
\end{figure}

\section{Connections to octahedral modes and dynamics}

\begin{figure}[p]
\centering\includegraphics[width=0.98\textwidth]{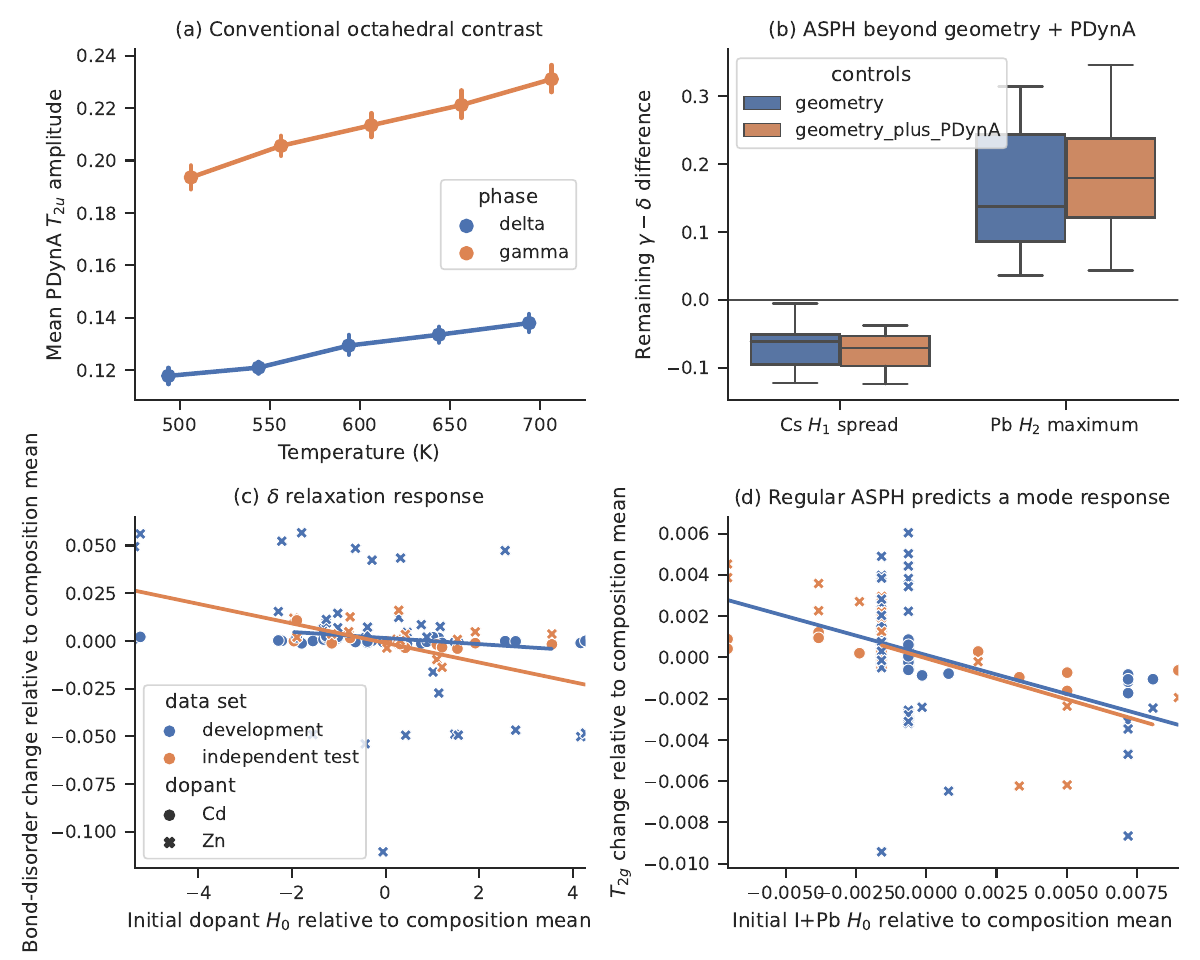}
\caption{\label{fig:s14}Connections between octahedral modes and topology. (a) The PDynA $T_{2u}$ amplitude provides a conventional local phase contrast. (b) Complete-network Cs $H_1$ and Pb $H_2$ retain a phase difference after subtracting predictions from either the 97-quantity geometry basis alone or that basis augmented by all PDynA mode amplitudes. (c,d) In substituted structures, initial dopant or host-network topology follows the change in bond-disorder spread and the $T_{2g}$ mode during DFT relaxation. Values are expressed relative to the mean at the same dopant identity and substitution count so that the comparison concerns arrangement rather than composition.}
\end{figure}

PDynA symmetry-mode amplitudes provide a local distortion basis for comparison with \ASPH\ (Figure~\ref{fig:s14}). Selected directed channels follow recognizable octahedral modes, while complete-sublattice Cs-loop heterogeneity and Pb-cavity persistence remain distinct. Symmetry modes identify how a well-defined octahedron distorts; topology additionally retains the organization among octahedra and remains available when an ideal-octahedron fit fails.

\begin{figure}[p]
\centering\includegraphics[width=0.98\textwidth]{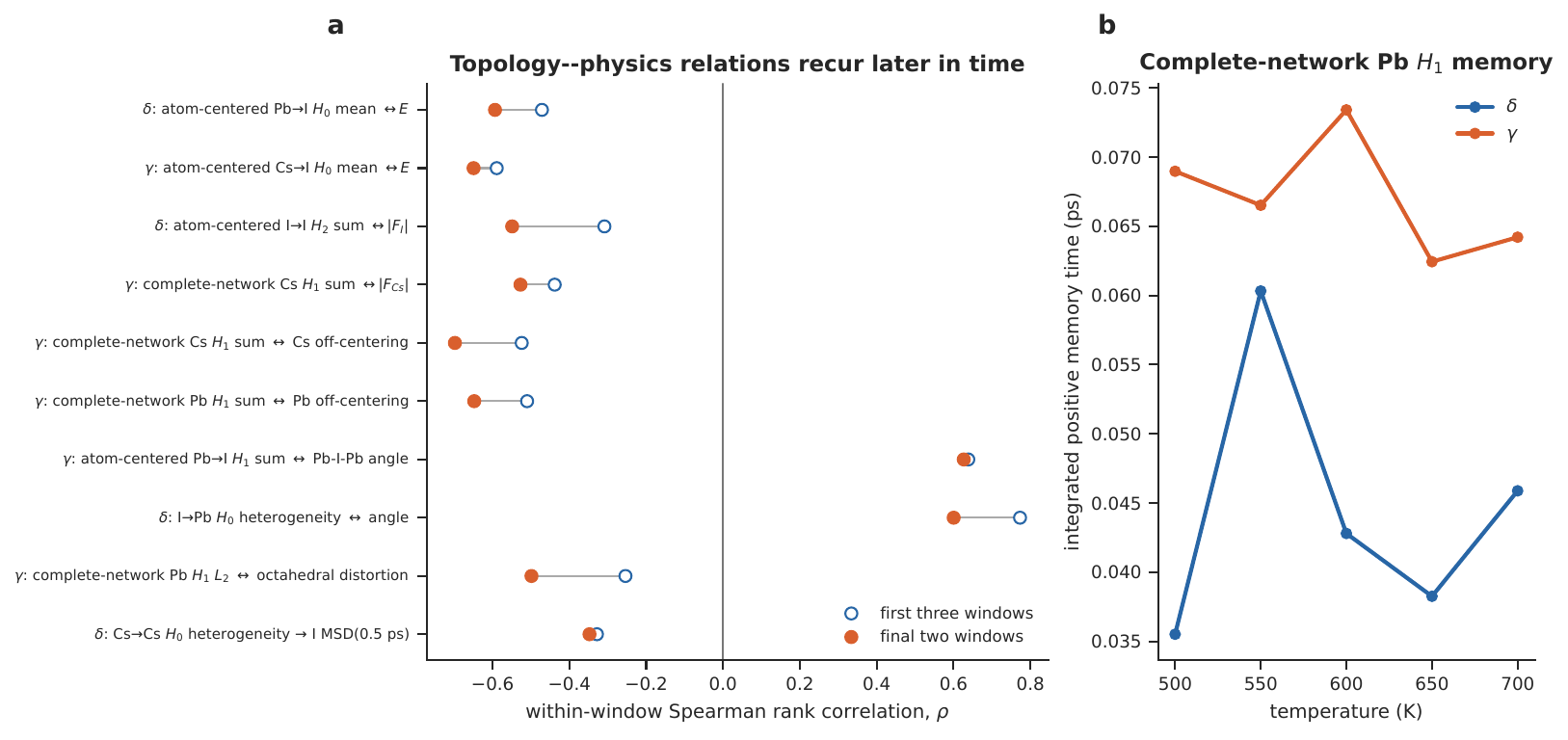}
\caption{\label{fig:s15}Topology--physics relationships across time. (a) Ten relationships identified in the first three separated windows recur with the same physical direction in the final two windows. Each line joins the median rank correlation used to define a relationship to the median obtained when that fixed relationship is evaluated in the later windows. (b) The integrated autocorrelation of complete-network Pb $H_1$ shows longer topological memory in $\gamma$ throughout 500--700 K.}
\end{figure}

Atom-centered and complete-sublattice \ASPH\ time series connect this organization to atomic dynamics (Figure~\ref{fig:s15}). Spearman's $\rho$ measures a monotonic rank association and ranges from $-1$ for opposite ordering to $+1$ for identical ordering. Complete-sublattice Cs $H_1$ decreases as Cs off-centering increases ($\rho\simeq-0.70$); complete-sublattice Pb $H_1$ similarly decreases with Pb off-centering in $\gamma$ ($\rho\simeq-0.65$); and Pb$\rightarrow$I loop persistence increases with the Pb--I--Pb angle ($\rho\simeq0.63$). We also compare a topological value at time $t$ with the iodide mean-square displacement accumulated after $t$. Greater heterogeneity of Cs$\rightarrow$Cs $H_0$ connectivity precedes smaller iodide displacement ($\rho\simeq-0.35$), linking a cage-network fluctuation to subsequent motion.

Iodide relaxation is measured by the self-intermediate scattering function
\begin{equation}
F_s(q,t)=\frac{1}{N_{\rm I}}\sum_{j=1}^{N_{\rm I}}\left\langle\exp\{i\mathbf q\cdot[\mathbf r_j(t_0+t)-\mathbf r_j(t_0)]\}\right\rangle_{t_0,\,|\mathbf q|=q}.
\end{equation}
It is the average memory of an iodide's initial position on length scale $2\pi/q$. The continuously unwrapped fractional trajectory is mapped into the time-mean cell and rigid center-of-mass translation is removed; homogeneous NPT cell breathing is therefore not counted as atomic displacement. The static structure factor is averaged over 32 fixed, approximately uniform directions and 24 frames on a 0.05 \AA$^{-1}$ grid. Its strongest peak in the prespecified 1.0--2.2 \AA$^{-1}$ interval fixes $q$ for each phase. $F_s$ uses the analytic orientational average $\sin(qr)/(qr)$ and time origins every 0.05 ps. The long-time plateau is the median from the final third of the available 30 ps lag window. After subtracting that plateau and normalizing the remaining amplitude to unity, the relaxation time is obtained by linear interpolation at the first $1/e$ crossing. The dynamic susceptibility is the atom-count-scaled variance across time origins of the corresponding single-particle overlap and measures how unevenly different iodides relax.

The Pb--I cage stiffness uses Pb displacement from the centroid of its six nearest iodides. For each crystallographic Pb site, the time-mean displacement and time-mean force are removed. A single isotropic restoring coefficient is then fitted to all Cartesian components by least squares, $\mathbf F=-k\mathbf u$, or equivalently $k=-\sum\mathbf u\cdot\mathbf F/\sum|\mathbf u|^2$; 1 ps blocks provide its standard error. Complete-network topological memory is calculated in each separated 1 ps window after linear detrending and removal of the instantaneous cell-length contribution. Its integrated positive autocorrelation time is $\Delta t[1/2+\sum_{n=1}^{n_0-1}C(n)]$, where $C(n)$ is the normalized correlation between values separated by $n$ saved frames, $n_0$ is the first lag at which that correlation is nonpositive, and $\Delta t=0.01$ ps. The quantity is the area under the initial positive part of the correlation curve and therefore measures how long a fluctuation retains memory.

\begin{figure}[p]
\centering\includegraphics[width=0.98\textwidth]{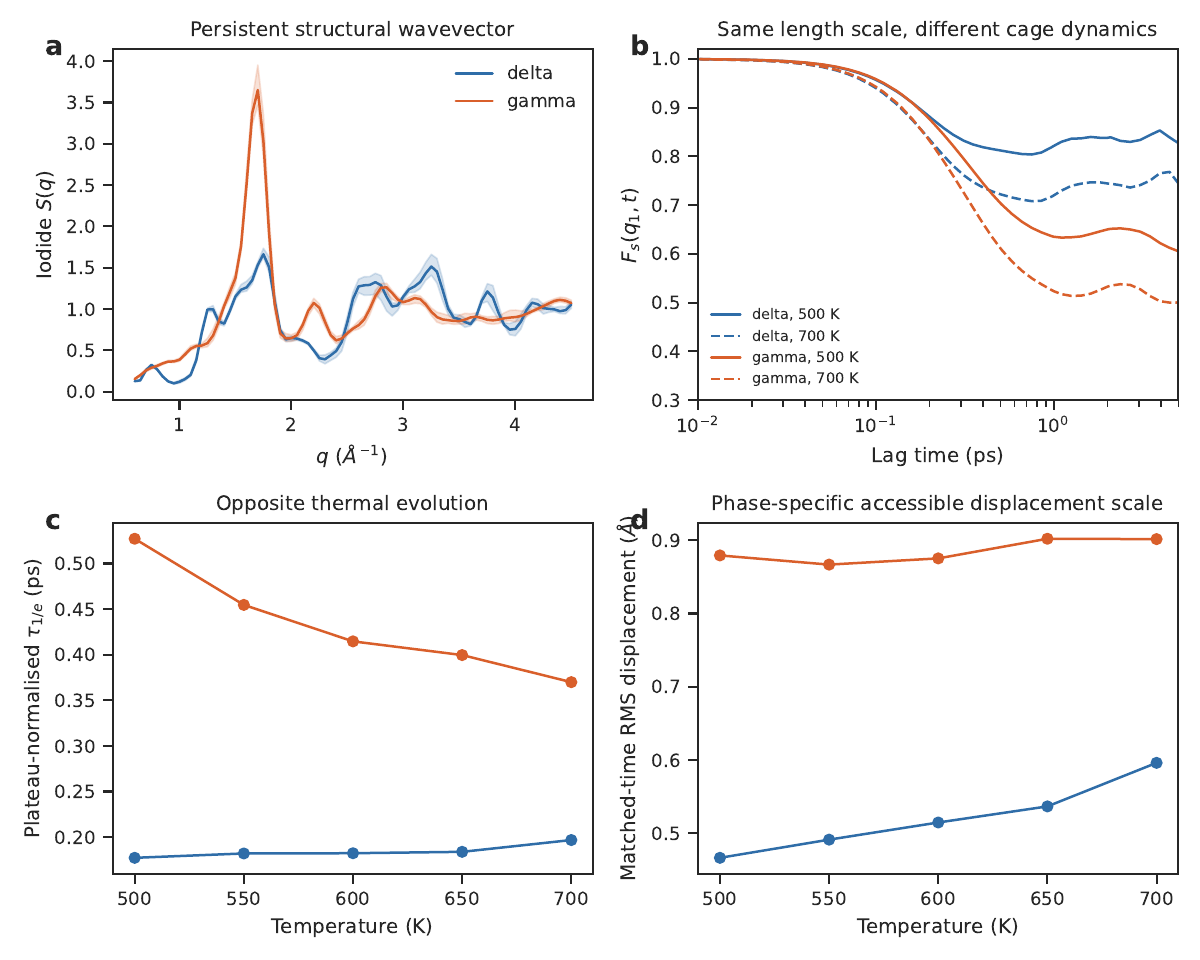}
\caption{\label{fig:s16}Separation of structural length and dynamical time scales. (a) Powder-averaged iodide structure factors identify the principal wavevectors used for each phase. (b) Plateau-normalized self-intermediate scattering functions at 500 and 700 K. (c) The $1/e$ self-relaxation time across temperature. (d) Root-mean-square iodide displacement reached at that phase- and temperature-specific relaxation time. The larger $\gamma$ displacement occurs within a caged, non-diffusive regime.}
\end{figure}

The primary peaks are $q=1.75$ \AA$^{-1}$ in $\delta$ and 1.70 \AA$^{-1}$ in $\gamma$, corresponding to nearly equal real-space lengths of 3.59 and 3.70 \AA\ (Figure~\ref{fig:s16}). Despite this similarity, the median of the five temperature-matched $\gamma/\delta$ relaxation-time ratios is 2.27 and remains between 2.13 and 2.29 when calculated separately in four nonoverlapping 16 ps production segments. From 500 to 700 K, the $\delta$ relaxation time changes from 0.18 to 0.20 ps, whereas the $\gamma$ value decreases from 0.53 to 0.37 ps. The lower $\gamma$ long-time plateau, its 5.2-fold larger maximum dynamic susceptibility, and the larger root-mean-square displacement reached at the $1/e$ time identify more extensive and more heterogeneous caged motion rather than long-range diffusion. Force--displacement slopes give Pb--I cage stiffnesses of 1.41--1.60 eV \AA$^{-2}$ in $\delta$ and 1.01--1.07 eV \AA$^{-2}$ in $\gamma$. The median of the five temperature-matched ratios of complete-sublattice Pb $H_1$ memory is 1.63; the phase medians themselves are 0.067 and 0.043 ps. The longer $\gamma$ memory connects its slower cage response to persistence of the corner-sharing Pb network. In $\delta$, this time is maximal at 550 K: the median over five separated windows is 0.060 ps, compared with 0.036--0.046 ps at 500, 600, 650, and 700 K, and approaches the $\gamma$ value of 0.067 ps. Individual-window values overlap across temperatures, so the excursion is interpreted together with the independent 550 K minimum in geometry-conditioned phase contrast rather than as a transition-temperature estimate.

\clearpage
\section{Dopant topology, relaxation, and model consistency}

The database of Eremin et al. contains 73,962 enumerated 160-atom cells in which Cd or Zn replaces Pb. DFT-relaxed energies are available for 202 of them. The lower-substitution reference set contains 142 structures: every enumerated structure with zero, one, or two substitutions and a sample with three substitutions. Descriptor choices, the number of principal components, and the regression penalty are determined only from this reference set. A separate test set contains 60 structures with three to five substitutions; the four- and five-substitution cells therefore test transfer to compositions absent from the reference set. Their DFT energies are used only after the descriptor choice, principal-component reduction, and regression penalty have been fixed. The remaining 73,760 structures have predictions from the supplied graph neural networks (GNNs) but no DFT labels and are used only to examine their enumerated energy landscapes. The published DFT calculations used VASP, projector-augmented-wave potentials, the Perdew--Burke--Ernzerhof (PBE) exchange--correlation functional, a 600 eV plane-wave cutoff, and sampling at the Brillouin-zone center ($\Gamma$ point). Every descriptor is calculated from the same initial, unrelaxed coordinates supplied to the neural networks.

For each cell, only the Cd or Zn sites are placed in a periodic point cloud. Initially every dopant is a separate $H_0$ component. As the connection distance increases, components merge; the finite merge distances are exactly the edge lengths of the periodic minimum-spanning connectivity hierarchy. Small distances indicate a compact dopant group, whereas large distances indicate dispersion across the Pb sublattice.

The common diagram-extraction routine returns 28 $H_0$ summaries. Three count the finite, positive-lifetime, and nonterminating components. For each of birth, death, and lifetime it records seven distribution statistics---minimum, mean, standard deviation, maximum, lower quartile, median, and upper quartile---giving $3\times7=21$ values. Total persistence and its Euclidean norm add two, and persistence entropy and its exponential add two more: $3+21+2+2=28$. In $H_0$, all components are born at zero and death equals lifetime, so some summaries are mathematically redundant. They are retained because the same extractor is used for $H_0$--$H_2$. In the primary ordering model, these 28 values are combined with the seven composition terms defined below. Each of the resulting 35 inputs is centered and divided by its standard deviation over the 142 lower-substitution reference structures. Principal-component analysis then replaces the correlated inputs by five mutually orthogonal linear combinations, the minimum number that retains 95\% of their total variation. Thus five fitted combinations of one connectivity hierarchy and composition enter the regression, rather than 28 separately asserted mechanisms. $H_1$ and $H_2$ contain few or no finite features at the substitution counts represented most strongly in the reference set and are not used for the primary ordering model.

Pair-ranking accuracy measures whether a descriptor selects the lower-energy arrangement at fixed chemistry. Structures are first grouped by phase, dopant identity, and substitution count. Every pair within a group is scored as correct when the predicted energy difference has the same sign as the DFT difference; ties receive half credit, and the resulting accuracy is averaged across groups. Random pair ordering gives 0.5. Composition alone also gives 0.5 because it assigns the same value to every arrangement within a group. For the 95\% interval, these chemical groups are resampled rather than their individual pairs; selecting a group retains all of its structures and pairwise comparisons, including pairs that share a structure.

Every regression receives the same seven composition terms: indicators for phase and dopant identity, substitution count, its square, and the phase--count, dopant--count, and phase--dopant interactions. These terms allow the mean energy to vary with chemistry but carry no information about arrangement within a fixed-composition set. The dopant-topology model adds the 28 uniformly extracted $H_0$ summaries described above. The SOAP comparator uses the periodic DScribe implementation with species Cs, Pb, I, Cd, and Zn, $r_{\rm cut}=6$ \AA, $n_{\max}=l_{\max}=4$, and inner averaging over Pb, Cd, and Zn B-site centers. It produces a 1050-component cell vector, giving $1050+7=1057$ inputs before principal-component reduction. Missing inputs are replaced by the median of that variable in the 142 lower-substitution structures; centering, scaling, principal components, and ridge penalties are also determined from those structures alone. To select the regression penalty, the 142 structures are divided four times into fitting and validation portions while keeping every fixed-composition set intact. The value with the smallest validation error is fixed, and the complete procedure is then applied unchanged to the 60 higher-substitution structures. On that evaluation set, the composition--$H_0$ model gives $R^2=0.925$ and a mean absolute error of 1.71 meV atom$^{-1}$, compared with 0.48 meV atom$^{-1}$ for the best supplied neural-network consensus. Its within-composition pair-ranking accuracy is 0.658, compared with 0.500 for composition alone, 0.392 for composition plus dopant-distance summaries, and 0.283 for composition plus SOAP.

The graph-neural-network consistency test isolates arrangement from composition before comparing structure and energy. Within each fixed-composition set, the mean energy is subtracted from every energy and the mean $H_0$ is subtracted from every $H_0$ value. This centering removes the offsets between phases, dopant species, and substitution counts while preserving how configurations are ordered within each set. Spearman correlation is then calculated over the centered values. A positive correlation means that, relative to other structures of the same composition, compact arrangements (small $H_0$) are lower in energy and dispersed arrangements are higher. A negative correlation reverses this ordering. The test therefore measures whether the model energy has the same dependence on dopant arrangement as the DFT energy.

The complete two-dopant spaces permit a direct statistical-mechanical calculation. They contain 15 or 16 symmetry-inequivalent structures for each phase and dopant, and the corresponding degeneracies sum to 496 microscopic arrangements. Formation energies reported per atom are multiplied by 160 to obtain the energy of the complete simulation cell. For inequivalent structure $i$, with cell energy $E_i$ and $g_i$ symmetry-equivalent realizations, the canonical probability is
\begin{equation}
p_i(T)=\frac{g_i\exp[-(E_i-E_{\min})/(k_{\rm B}T)]}{\sum_j g_j\exp[-(E_j-E_{\min})/(k_{\rm B}T)]}.
\end{equation}
This is the standard Boltzmann weighting: low-energy structures receive greater probability, while $g_i$ accounts for how many microscopic arrangements each representative structure denotes. Eremin et al. used $k_{\rm B}T=0.0257$ eV, corresponding approximately to 300 K, for their Boltzmann-averaged configurational energies. We retain that temperature for the topology-resolved population analysis, using the exact value $k_{\rm B}T$ with $T=300$ K. The calculation uses static DFT energies and is unrelated to the 500--700 K AIMD trajectories. Results at 500 and 700 K test how the inferred selection changes with temperature.

An inequivalent class $i$ represents $g_i$ microscopic arrangements, each with probability $p_i/g_i$. The microscopic configurational entropy is therefore
\begin{equation}
\frac{S_{\rm conf}}{k_{\rm B}}=-\sum_i p_i\ln\!\left(\frac{p_i}{g_i}\right),
\end{equation}
and $N_{\rm eff}=\exp(S_{\rm conf}/k_{\rm B})$ is the number of equally probable microscopic arrangements with the same entropy. At 300 K, $N_{\rm eff}$ is 73 for $\delta$--Zn and 468 for $\delta$--Cd out of 496 possible arrangements; their most probable inequivalent classes carry 76\% and 15\% of the probability, respectively. Energy weighting changes the mean dopant $H_0$ relative to weighting by degeneracy alone by $-4.30$ \AA\ for $\delta$--Zn but only $-0.40$ \AA\ for $\delta$--Cd. The negative sign denotes selection toward more compact dopant arrangements. The same conclusion persists at 500 and 700 K: $\delta$--Zn retains 137 and 187 effective arrangements with $H_0$ shifts of $-3.22$ and $-2.54$ \AA, whereas $\delta$--Cd retains 487 and 492 with shifts of $-0.19$ and $-0.12$ \AA.

The sign inversion has a direct thermodynamic meaning. Within the complete static two-substitution spaces, DFT selects smaller $H_0$ in all four phase--dopant combinations. The GNN-only configurations span 60 thermodynamic conditions: two phases, two dopants, five substitution counts, and three temperatures. Applying each of the six Allegro family means to these conditions gives $6\times60=360$ family--condition results; 300 have a negative thermal $H_0$ shift and therefore favor compact arrangements. The two SchNet family means give $2\times60=120$ results, all with positive shifts of approximately $+0.02$ to $+0.31$ \AA\ and therefore favoring more dispersed arrangements. SchNet would consequently increase the population of dispersed configurations where the available DFT data select clustering, changing the dominant motifs and configurational entropy. The supplied predictions comprise eight model families---two training scopes crossed with SchNet or Allegro and, for Allegro, three pretraining choices---with 48 fitted models in each family, for $8\times48=384$ models. In the supplied identifiers, \texttt{both\_both} means that one family is trained jointly on Cd- and Zn-doped structures from both phases, whereas \texttt{element\_both} uses a separate family for each dopant while retaining both phases. The suffixes \texttt{non-pr}, \texttt{ocpr}, and \texttt{aflowpr} denote no pretraining, Open Catalyst Project pretraining, and AFLOW pretraining, respectively. Among the 247 models whose higher-substitution mean absolute error is below 1 meV atom$^{-1}$, 61 nevertheless give the wrong sign for the centered $H_0$--energy relation.

The atom-centered calculation uses an 8.5 \AA\ periodic neighborhood and identifies where the host lattice responds to the dopant arrangement. The smallest half lattice translation among all 73,962 initial 160-atom cells is 8.5766 \AA; 8.5 \AA\ is therefore the largest one-decimal radius that admits a unique minimum-image neighborhood throughout the database. This is the same construction rule that gives 9.5 \AA\ for the larger AIMD cells. In each $\delta$ structure, the rank dependence of both the initial I$\rightarrow$Pb $H_0$ first-connection scale and subsequent iodide displacement on nearest-dopant distance, number of neighbors, and point-cloud radius is first removed. The correlation between the remaining quantities is a partial rank correlation: it asks whether topology remains related to relaxation after accounting for these simpler local-size and distance measures. Its median is 0.29 across the 70 lower-substitution $\delta$ structures used to identify the relation and 0.38 across the 30 higher-substitution $\delta$ test structures, with a positive sign in all 30. A larger first connection scale identifies an iodide that is initially less tightly integrated into its local Pb framework. Two maps of structural change locate the accompanying reorganization: in $\gamma$, Cs displacement follows the relaxation-induced change in Cs$\rightarrow$Pb $H_0$ connectivity (median $\rho=0.73$ in the lower-substitution structures and 0.65 in the higher-substitution structures); in $\delta$, iodide displacement follows the change in I$\rightarrow$I $H_2$ persistence ($\rho=0.38$ and 0.42). All directed channels and $H_0$--$H_2$ summaries are first evaluated in the 142 lower-substitution structures. A relationship is reported only if its whole-structure Wilcoxon test remains significant after Benjamini--Hochberg correction within each phase and analysis type, retains the same direction in the 60 higher-substitution structures, and passes the same correction there. The I$\rightarrow$Pb relationship is prospective because it uses only the initial coordinates; the two response maps describe where reorganization occurred.

\begin{table}[h]
\centering
\caption{\label{tab:doped}Doped-data partitions and their role in the analysis.}
\begin{tabular}{lrrl}
\toprule
Partition & structures & atoms/cell & use\\
\midrule
DFT lower-substitution set & 142 & 160 & descriptor and regression choices\\
DFT higher-substitution set & 60 & 160 & evaluation after all choices are fixed\\
GNN-only configurations & 73,760 & 160 & thermal ordering across enumerated landscapes\\
\bottomrule
\end{tabular}
\end{table}

\begin{table}[h]
\centering
\caption{\label{tab:gnn}Energy error and arrangement--energy relation on the 60 higher-substitution structures for every supplied model family. Each neural-network row is the mean prediction of 48 fitted models. MAE is the mean absolute energy error in meV atom$^{-1}$. The rank coefficient $\rho(H_0,E)$ is calculated after the separate mean of each fixed-composition set has been removed from both energy and the mean finite $H_0$ connection scale; positive values preserve the DFT preference for compact arrangements. The supplied family identifiers are retained for reproducibility. Figure 6c compares the two matched \texttt{both\_both}, non-pretrained families, holding training scope and pretraining fixed.}
\begin{tabular}{lrr}
\toprule
Method or supplied family identifier & MAE & $\rho(H_0,E)$\\
\midrule
DFT reference & --- & +0.49\\
\texttt{both\_both\_schnet\_non-pr} & 0.61 & $-0.78$\\
\texttt{element\_both\_schnet\_non-pr} & 1.31 & $-0.81$\\
\texttt{both\_both\_allegro\_non-pr} & 0.66 & +0.72\\
\texttt{both\_both\_allegro\_aflowpr} & 0.71 & +0.24\\
\texttt{both\_both\_allegro\_ocpr} & 0.78 & +0.74\\
\texttt{element\_both\_allegro\_non-pr} & 0.55 & $-0.03$\\
\texttt{element\_both\_allegro\_aflowpr} & 11.82 & +0.25\\
\texttt{element\_both\_allegro\_ocpr} & 2.16 & +0.74\\
\bottomrule
\end{tabular}
\end{table}

\begin{figure}[p]
\centering\includegraphics[width=0.98\textwidth]{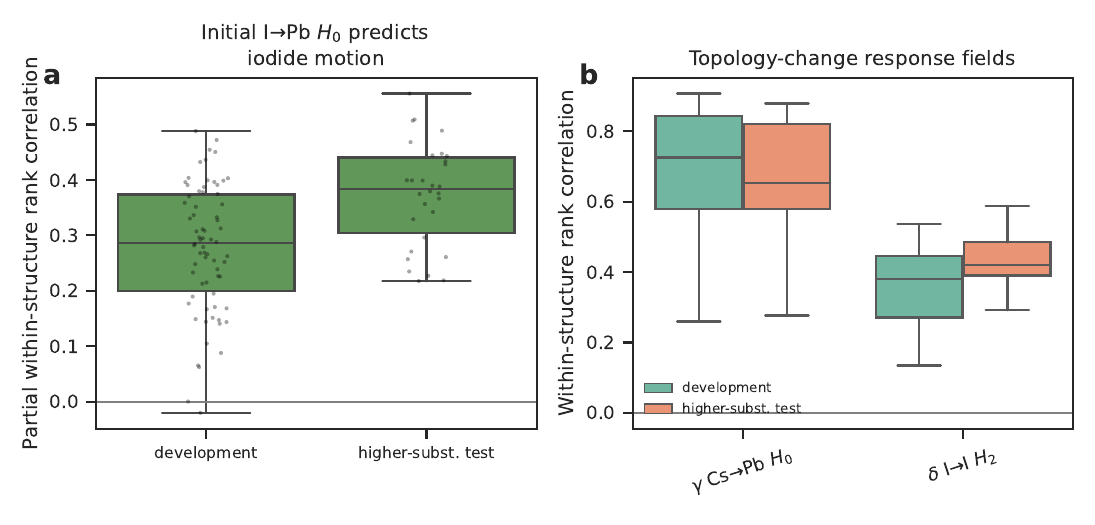}
\caption{\label{fig:s17}Local and spatial response in substituted structures. (a) The initial $\delta$-phase I$\rightarrow$Pb $H_0$ first-connection scale is associated with subsequent iodide displacement after removing the effects of nearest-dopant distance, number of points, and cloud radius (test partial $\rho=0.38$, positive in all 30 test structures). (b) Directed response maps show that Cs$\rightarrow$Pb $H_0$ reorganization accompanies Cs displacement in $\gamma$, while I$\rightarrow$I $H_2$ reorganization accompanies iodide displacement in $\delta$.}
\end{figure}

\bibliographystyle{unsrtnat}
\bibliography{references}